\documentclass[sigconf,nonacm]{acmart}

\usepackage{amsmath,amsfonts}
\usepackage{stmaryrd}
\usepackage{mathtools}
\usepackage{mathbbol}
\usepackage{soul}
\usepackage{tikz}
\usetikzlibrary{arrows,shapes,fadings,positioning,patterns,decorations,calc,decorations.pathreplacing,calligraphy}
\usepackage{xspace}
\usepackage{semantic}

\theoremstyle{acmplain}

\theoremstyle{acmdefinition}

\usepackage{algorithm} 
\usepackage[noend]{algpseudocode}

\usepackage{tabularx}

\usepackage{eurosym}

\usepackage{listings}
\newcommand{\kwcol}{black}
\lstdefinestyle{sql}{%
  language=SQL,%
  basicstyle=\fontfamily{pcr}\small\fontsize{7.5}{8}\selectfont,%
  keywordstyle=\bfseries\color{\kwcol},%
  morekeywords={RETURN,RETURNS,EXECUTE,QUERY,REPLACE,FUNCTION,LANGUAGE},%
  deletekeywords={YEAR},
  literate={\\\$}{\\\$}1 {\\\%}{\\\%}1
}
\lstnewenvironment{sql}[1][]{%
  \renewcommand{\kwcol}{violet}
  \lstset{style=sql,#1}%
}{}

\usepackage[shortlabels]{enumitem}
\setlist{labelindent=\parindent, leftmargin=*}

\usepackage{subcaption}

\usepackage{framed}

\copyrightyear{2025}
\acmYear{2025}
\setcopyright{cc}
\setcopyright{cc}
\setcctype{by}

\acmConference[]{}{}{}
\acmBooktitle{}
\acmDOI{}
\acmISBN{}

\newcommand{\OMIT}[1]{}

\newcommand{\truthvalue}[1]{\mbox{\big\langle\!\!\big\langle}{#1}\mbox{\big\rangle\!\!\big\rangle}}

\usepackage{listings}

\usepackage{bold-extra}

\makeatletter

\newcommand{\setst}{\@ifstar{\autosetst}{\paramsetst}}
\newcommand{\autosetst}[2]{\left\lbrace\,#1~\middle|~#2\,\right\rbrace}
\newcommand{\paramsetst}[3][]{#1\{\,#2\mathbin{#1|}#3\,#1\}}
\newcommand{\bag}{\@ifstar{\autobag}{\parambag}}
\newcommand{\autobag}[1]{\llbrace*#1\rrbrace*}
\newcommand{\parambag}[2][]{\llbrace[#1]#2\rrbrace[#1]}
\newcommand{\bagst}{\@ifstar{\autobagst}{\parambagst}}
\newcommand{\autobagst}[2]{\llbrace*\,#1\,\middle|\,#2\,\rrbrace*}
\newcommand{\parambagst}[3][]{\llbrace[#1]\,#2\mathbin{#1|}#3\,\rrbrace[#1]}

\newcommand{\llbrace}{\@ifstar{\leftllbrace}{\paramllbrace}}
\newcommand{\paramllbrace}[1][]{{#1\{\hspace*{-.25em}#1\{}}
\newcommand{\leftllbrace}{\left\lbrace\kern-3\nulldelimiterspace\middle\lbrace}
\newcommand{\rrbrace}{\@ifstar{\rightrrbrace}{\paramrrbrace}}
\newcommand{\paramrrbrace}[1][]{{#1\}}\hspace*{-.25em}{#1\}}}
\newcommand{\rightrrbrace}{\middle\rbrace\kern-3\nulldelimiterspace\right\rbrace}
\makeatother

\newcommand{\gqlil}[1]{\lstinline`#1`}

\newcommand{\dom}{\textsf{dom}}

\newcommand{\sem}[1]{\left\llbracket#1\right\rrbracket}

\tikzset{>=stealth}
\tikzset{
  detail/.style={%
    draw, rectangle split, rectangle split parts=#1,
    rectangle split draw splits=true
  },
  onslide/.code args={<#1>#2}{\only<#1>{\pgfkeysalso{#2}}},
  invisible/.style={opacity=0},
  visible on/.style={alt=#1{}{invisible}},
  alt/.code args={<#1>#2#3}{%
      \alt<#1>{\pgfkeysalso{#2}}{\pgfkeysalso{#3}} %
  },
}
\tikzset{%
  nodeid/.style={%
    circle, draw, minimum size=4mm, line width=0.5mm, inner sep=1mm, outer sep=0.75mm,
  },
  edgeid/.style={%
    line width=0.5mm, circle, minimum size=5mm, draw=none, fill=white, inner sep=0mm, outer sep=1mm,
  },
  relationship/.style={%
    draw, very thick, -latex
  },
  newly created/.style={%
  },
  existing/.style={%
    draw=gray,
  },
  edgelabel/.style={%
    pos=.433,
    above,
  }
}

\makeatletter
\tikzset{
    dot diameter/.store in=\dot@diameter,
    dot diameter=3pt,
    dot spacing/.store in=\dot@spacing,
    dot spacing=10pt,
    dots/.style={
        line width=\dot@diameter,
        line cap=round,
        dash pattern=on 0pt off \dot@spacing
    }
}
\makeatother

\usepackage{color}
\definecolor{gray}{rgb}{0.4,0.4,0.4}
\definecolor{darkblue}{rgb}{0.0,0.0,0.6}
\definecolor{darkred}{rgb}{0.45,0,0}
\definecolor{darkgreen}{rgb}{0,0.30,0.20}
\definecolor{darkpurple}{RGB}{120, 0, 180}
\colorlet{keywordcolor}{darkblue}
\newcommand{\kwfont}{\color{keywordcolor}\bfseries}

\colorlet{labelcolor}{darkgreen}
\newcommand{\lblfont}{\color{labelcolor}}

\colorlet{keycolor}{darkred}
\newcommand{\keyfont}{\color{keycolor}}

\colorlet{structcolor}{black}
\newcommand{\structfont}{\color{structcolor}\bfseries}

\usepackage{listings,textcomp}
\lstdefinelanguage{cypher}
{
  columns=fullflexible,
  otherkeywords={<,>,-,[,],\{,\},|,:,?,*},
  keywords=[1]{},
  keywordstyle=,
  keywordstyle=[1]\color{darkpurple},
  keywords=[2]{CONTINUE, ACYCLIC, SIMPLE, MATCH,KEEP,WHERE,WITH,OPTIONAL,RETURN,MERGE,CREATE,SET,DETACH,DELETE,REMOVE,ALL,ANY,GROUP,SAME,IS,NOT,AND,THEN,OR,YIELDS,DISTINCT,PATHS,SHORTEST,CHEAPEST,TRAIL,WALK,COST},
  keywordstyle=[2]\kwfont,
  keywords=[3]{<,>,-,[,],(,)},
  keywordstyle=[3]\structfont,
  keywords=[5]{Woman,L1,L2,L3,L3},
  keywordstyle=[5]\lblfont,
  keywords=[6]{BOUGHT_WITH,MOVIE,FRIEND,PARTNER,BOOK,TRANSFERS,CORP,BANK,SPOUSE,PERSON},
  keywordstyle=[6]\lblfont,
  keywords=[7]{name,id,genre},
  keywordstyle=[7]\keyfont,
  string=[m]{"},
  stringstyle=,
}
\lstnewenvironment{cypher}[1][]{\lstset{language=cypher,#1}}{}

\newcommand{\forcedhfill}{\hspace*{0pt plus 1fill}}
\makeatletter
\newcounter{query}
\renewcommand{\thequery}{(\arabic{query})}
\newcounter{nextquery}

\newlength{\lstskip}
\newcommand{\vm@ccypher@start}[2][]{%
  \leavevmode\unskip\pagebreak[1]\vspace{\lstskip}\par\noindent%
  #1%
  \lst@boxtrue%
  \lstset{language=cypher,boxpos=b,resetmargins=true,mathescape=true,#2}%
  \forcedhfill}
\newcommand{\vm@ccypher@end}[1][]{\forcedhfill#1\par\addvspace{\lstskip}}
\lstnewenvironment{gql}[1][]%
  {\vm@ccypher@start[\refstepcounter{query}]{#1}}%
  {\vm@ccypher@end[~\thequery]}
\lstnewenvironment{gql*}[1][]%
  {\vm@ccypher@start{#1}}
  {\vm@ccypher@end}
\makeatother

\newcommand{\arclit}[1]{%
  \if\relax\detokenize{#1}\relax
  \mathrel{\smash{\xLeftrightarrow{}}}
  \else
  \mathrel{\xLeftrightarrow{\,#1\,}}
  \fi
}
\newcommand{\rightlit}[1]{%
  \if\relax\detokenize{#1}\relax
  \mathrel{\smash{\xrightarrow{}}}
  \else
  \mathrel{\xrightarrow{\,#1\,}}
  \fi
}
\newcommand{\leftlit}[1]{%
  \if\relax\detokenize{#1}\relax
  \mathrel{\smash{\xleftarrow{}}}
  \else
  \mathrel{\xleftarrow{\,#1\,}}
  \fi
}

\newcommand{\undirlit}[2][-\hspace{-3pt}-\hspace{-3pt}-]{%
  \if\relax\detokenize{#2}\relax
  \mathrel{#1}
  \else
  \stackrel{#2}{#1}
  \fi
}

\newcommand{\const}{\ensuremath{\textsf{Const}}}

\newcommand{\emptytup}{\tup{}}

\def\stoptoken{!}
\newif\ifpathstop
\def\vmedge#1,{\if\stoptoken#1\let\vmnode\relax\else,\textcolor{red}{#1},\fi\vmnode}
\def\vmnode#1,{\if\stoptoken#1\let\vmedge\relax\else\textcolor{blue}{#1}\fi\vmedge}%

\definecolor{darkblue}{RGB}{0,56,153}
\definecolor{darkred}{RGB}{153,0,0}

\newcommand{\Sem}[1]{\ensuremath{\left\llbracket{#1}\right\rrbracket}}

\renewcommand{\emptyset}{\varnothing}
\newcommand{\sexpr}[1]{{\normalfont\textit{#1}}}
\newcommand{\bexpr}{\textit{Bindings}}

\newcommand{\rexpr}{\textit{Expr}}
\newcommand{\formula}{\textit{Formula}}
\newcommand{\argr}{\textit{Arg}}

\newcommand{\ddd}[1]{#1\raisebox{.4ex}{...}}

\newcommand{\Values}{\textbf{Values}}
\newcommand{\Ids}{\textbf{IDs}}
\newcommand{\VarIds}{\ddd{\Ids}}
\newcommand{\Tuples}{\textbf{Tuples}}
\newcommand{\Rels}{\textbf{Rels}}

\newcommand{\powerset}{\mathcal{P}}

\newcommand{\venv}[1][\mu]{{\ensuremath{#1}}}

\newcommand{\renv}[1][\mathcal{R}]{{\ensuremath{#1}}}
\newcommand{\rev}[1]{{\ensuremath{\mathbin{\stackrel{#1}{\oplus}}}}}

\newcommand{\lov}{\rev{}}

\newcommand{\relnames}{\mathbf{N}}

\renewcommand{\renv}{\mu}

\newcommand{\df}{:=}
\newcommand{\tup}[1]{\left\langle #1 \right\rangle}

\newcommand{\kw}[1]{\texttt{%
    \fontfamily{pcr}\bfseries\color{darkblue}\selectfont #1}}

\newcommand{\gr}[1]{\texttt{#1}}

\renewcommand{\renv}{\mu}

\newcommand{\nat}{\ensuremath{\mathbb{N}}}

\newcommand{\GNF}{GNF\xspace}
\newcommand{\GNFfull}{graph normal form\xspace}

\colorlet{varcolor}{darkpurple}
\newcommand{\varfont}{\color{varcolor}}
\colorlet{stringcolor}{black}
\newcommand{\stringfont}{\color{stringcolor}}

\usepackage{xspace}

\def\fixeddepthunderline#1{\underline{\rule[-.5ex]{0pt}{1pt}#1}}
\lstdefinelanguage{rel}
{
  columns=fullflexible,
  otherkeywords={.,\{,\},^},
  keywords=[1]{},
  keywordstyle=,
  keywordstyle=[1]\structfont,
  keywords=[2]{def,where,exists,forall,and,not,or,implies,iff,xor,reduce, in, namespace, end, ic, value, type, entity, requires, empty},
  keywordstyle=[2]\kwfont,
  keywords=[3]{RName, OrderWithPayment, PaymentOrder, OrderedProducts, PaymentAmount, Payment,OrderProductTotal,OrderTotal,OrderPaymentAmount,OrderPaid,OrderPayment,OrderProductQuantity,ClosedOrders, ProductPrice, OrderedProductPrice, OrderPayed, NotOrdered, NotP1Price, DiscountedproductPrice, add, modulo, AdditiveInverse, Int, PsychologicallyPriced, TC_E, ProductRS, ID, Perm, Prefix, Product, Union, Minus, Proj13, Select, Cond12, SameOrder, SameOrderDiffProduct, Expensive, BoughtWithExpensiveProduct, AllOrders, OutstandingOrders, total_pay, MatrixVector, ScalarProd, AlwaysOrdered, Ord, dimension, vector, range, abs, delta, next, stop, PageRank,DiscountedProductPrice},
  keywordstyle=[3]\lblfont,
  keywords=[4]{TC,Argmin,Argmax,E,E1,E2,sum,union,count,positive,valid_item,MatrixMult,mean,min,minimum,maximum,max,avg,pjoin,pjoinUV,dot_join,left_override,log,rel_primitive_add,rel_primitive_mult,rel_primitive_log,addUp,TrCl_E,prefix_join,rel_primitive_multiply,multiply,APSP},
  keywordstyle=[4]\lblfont,
  keywords=[5]{output,insert,delete},
  keywordstyle=[5]\lblfont\bfseries,
  keywords=[6]{i,j,k,x,y,z,x1,x2},
  keywordstyle=[6],
  keywords=[7]{A,B,U,V,E,F,R,S,Cond,G,P,M,Vec1,Vec2,Matrix},
  keywordstyle=[7]\varfont,
  keywords=[8]{scope1,scope2,std,common},
  keywordstyle=[8]\lblfont,
  string=[m]{"},
  stringstyle=\stringfont,
  mathescape=true,
  moredelim=*[is][\fixeddepthunderline]{__}{__},
}
\lstnewenvironment{rel}[1][]{\lstset{language=rel,#1}}{}

\makeatletter

\newcommand{\vm@crel@start}[2][]{%

  \normalfont%
  \pagebreak[0]\vspace{0pt plus 1pt}\par\noindent%
  #1%
  \lst@boxtrue%
  \lstset{language=rel,boxpos=b,#2}%
  \hfill}
\renewcommand{\forcedhfill}{\hspace*{0cm plus 1 fill}}
\newcommand{\vm@crel@end}[1][]{\forcedhfill#1\pagebreak[0]\vspace{.5ex plus 1pt}\par}
\lstnewenvironment{centeredrel}[1][]%
  {\vm@crel@start[\refstepcounter{equation}]{#1}}%
  {\vm@crel@end[\tagform@{\theequation}]}
\lstnewenvironment{centeredrel*}[1][]%
  {\vm@crel@start{#1}}
  {\vm@crel@end}
\makeatother

\newcommand{\relinline}[1]{\lstinline$#1$}

\let\relverb\lstinline

\newcommand{\Rel}{\textsf{Rel}\xspace}

\authorsaddresses{} %

\begin{document}

\title{Rel: A Programming Language for Relational Data}

\author{Molham Aref}
\affiliation{
    \institution{RelationalAI}
    \country{Woodside, USA}
}
\orcid{0009-0005-6584-5826}
\email{molham.aref@relational.ai}

\author{Paolo Guagliardo}
\affiliation{%
  \institution{University of Edinburgh}
    \country{Edinburgh, UK}
}
\orcid{0000-0003-0756-5787}
\email{paolo.guagliardo@ed.ac.uk}

\author{George Kastrinis}
\affiliation{
    \institution{RelationalAI}
    \country{Athens, Greece}
}
\orcid{0009-0002-5675-754X}
\email{george.kastrinis@relational.ai}

\author{Leonid Libkin}
\affiliation{
    \institution{RelationalAI \& Univ of Edinburgh}
    \country{Paris \& Edinburgh, France \& UK}
}
\orcid{0000-0002-6698-2735}
\email{leonid.libkin@relational.ai}

\author{Victor Marsault}
\affiliation{
    \institution{LIGM, Univ.~Gustave Eiffel, CNRS}
    \country{Marne-la-Vall\'ee, France}
}
\orcid{0000-0002-2325-6004}
\email{victor.marsault@univ-eiffel.fr}

\author{Wim Martens}
\affiliation{
    \institution{RelationalAI \& University of Bayreuth}
    \country{Bayreuth, Germany}
}
\orcid{0000-0001-9480-3522}
\email{wim.martens@relational.ai}

\author{Mary McGrath}
\affiliation{
    \institution{RelationalAI}
    \country{Little Compton, USA}
}
\orcid{0000-0003-3912-0117}
\email{mary.mcgrath@skylight.digital}

\author{Filip Murlak}
\affiliation{
    \institution{University of Warsaw}
    \country{Warsaw, Poland}
}
\orcid{0000-0003-0989-3717}
\email{f.murlak@uw.edu.pl}

\author{Nathaniel Nystrom}
\affiliation{
    \institution{RelationalAI}
    \country{Lugano, Switzerland}
}
\orcid{0009-0003-4405-0907}
\email{nate.nystrom@relational.ai}

\author{Liat Peterfreund}
\affiliation{%
  \institution{Hebrew University}
    \country{Jerusalem, Israel}
}
\orcid{0000-0002-4788-0944}
\email{liat.peterfreund@mail.huji.ac.il}

\author{Allison Rogers}
\affiliation{
    \institution{RelationalAI}
    \country{Seattle, USA}
}
\orcid{0000-0001-8557-7295}
\email{allison.rogers@relational.ai}

\author{Cristina Sirangelo}
\affiliation{%
  \institution{Universit\'e Paris Cit\'e, CNRS, IRIF}
    \country{ Paris, France}
}
\orcid{0000-0003-2559-512X}
\email{cristina@irif.fr}

\author{Domagoj Vrgo\v{c}}
\affiliation{
    \institution{PUC Chile}
    \country{Santiago, Chile}
}
\orcid{0000-0001-5854-2652}
\email{vrdomagoj@uc.cl}

\author{David Zhao}
\affiliation{
    \institution{RelationalAI}
    \country{Sydney, Australia}
}
\orcid{0000-0002-3857-5016}
\email{david.zhao@relational.ai}

\author{Abdul Zreika}
\affiliation{
    \institution{RelationalAI}
    \country{Melbourne, Australia}
}
\orcid{0000-0001-8812-5067}
\email{abdul.zreika@relational.ai}

\renewcommand{\shortauthors}{Molham Aref et al.}

\renewcommand{\shortauthors}{Molham Aref et al.}

\begin{abstract}
    From the moment of their inception, languages for relational data have been described as \emph{sublanguages} embedded in a host programming language. \Rel is a new relational language whose key design goal is to go beyond this paradigm with features that allow for programming in the large, making it possible to fully describe end to end application semantics. 
With the new approach  we can model the semantics of entire enterprise applications relationally, which helps significantly reduce architecture complexity and avoid the well-known \emph{impedance mismatch} problem. This paradigm shift is enabled by 50 years of database  research, making it possible to revisit the sublanguage/host language paradigm, starting from the fundamental principles. 
We present the main features of \Rel:  those that give it the power to express traditional query language operations and those that are designed to grow the language and allow programming in the large.

\end{abstract}



\maketitle

\section{Introduction}
\label{s:introduction}

From the moment of their inception, languages for relational data have been described as \emph{sublanguages} \cite{Codd71,Codd72}. This means that such languages are not designed to describe the entire end-to-end application logic of programs that involve data, but rather focus on specific operations that concern the storage, retrieval, or manipulation of data within such programs.
This view of database languages made a lot of sense fifty years ago, when database query languages were being introduced in an environment of procedural imperative programming, when declarative languages were a revolutionary idea, and when it was unclear if this idea could be extended beyond its domain-specific use.
Today, SQL is the prime success story of a declarative and domain-specific language. It remains in high demand %
by tech  employers \cite{top-pl}, %
which is a remarkable achievement for a 40-year-old programming (sub)language. 

The sublanguage paradigm however has important limitations. To start with, it entails having to communicate with a host programming language. This causes the well-known \emph{impedance mismatch}: the two languages have different data types, different data structures, different memory models, etc. Often they are even based on \emph{different programming paradigms}, one being fundamentally declarative and the other fundamentally imperative.  Furthermore, the two languages typically have separate runtime environments, which limits the optimizations that can be applied to programs that use both languages. The host language runtime environment may not support  features of the database such as automatic out of core computation, automatic parallelism, incremental computation.
In other words, the impedance mismatch 
\begin{enumerate}[nosep]
    \item causes extra complexity for developers, unrelated to solving the problem that they are actually trying to solve; and
    \item limits the capabilities of automatic support that we are used to in databases.
\end{enumerate}

Another effect of the sublanguage paradigm is that query languages are not equipped with important features needed for programming in the large, as these are delegated to the host language. Most notably, query languages lack support for building libraries. Indeed, SQL does not have a standard library, and when new features are needed they are added by means of expanding the language itself. Over the years, this led to an 11-part %
ISO standard comprising well over 4000 pages which is, for comparison, an order of magnitude larger than the C standard. This breaks a fundamental principle of programming language design formulated by Steele~\cite{Steele99}: 
\emph{define a small core and provide the functionality to build libraries}.

\Rel aims to go beyond the sublanguage paradigm. The database industry and research communities have accumulated a vast array of insights 
that reshuffle the cards on which the sublanguage design decision was based. We ask ourselves, can we \emph{design and implement} a language that natively handles data and semantics (or ``meaning'', as Codd called it in his Turing Award lecture \cite{Codd82}) in a database, preserving all the bedrock principles of databases (such as the data independence, communicability, and set-at-a-time processing objectives \cite{Codd82}),  %
providing the programmer with the necessary constructs to factor semantics out of application programs?
Such a full-featured language for data and semantics would allow for new powerful simplifications and optimizations owing to a single runtime environment, increasing the productivity of users and application developers. 

Towards this goal, \Rel has been designed and implemented 
as a  programming language for relational data that does away with the sublanguage paradigm.  Its key features are:  
\begin{enumerate}
    \item manipulation of both logical formulas and entire relations;
    \item powerful {\em recursion} built on the foundations of Datalog;
    \item \emph{abstraction} and \emph{application} as key constructs;
    \item variables that can \emph{range over tuples and relations}.
\end{enumerate}
We will touch upon all these points in the course of the paper.

A relational language beyond the sublanguage paradigm that supports programming in the large makes it possible to build database engines that can automate much more of the work being done by applications programmers working in the ``two language paradigm''. Building an engine for such a language is an ambitious goal that will take time to achieve. We believe, however, that it will bring significant gains over the sublanguage approach and that the time is ripe to embark on this journey.

\subsection*{A Rel Teaser}
The starting point of \Rel is Datalog with first-order logic formulas and aggregation in the bodies, which allows it to naturally express (recursive) database queries. We offer a few teasers to provide a glimpse of how the language goes beyond classical database querying and refer to Sections~\ref{sec:basics}--\ref{sec:programming-in-the-large} for deeper explanation. 

First, we define matrix multiplication as a general operation on relations.
Since relations can easily model vectors, matrices, and tensors, \Rel can naturally deal with analytics and ML workloads. 
Given an $n\times m$ matrix $\mathbf{A}$ and an $m\times p$ matrix $\mathbf{B}$, their product is an $n\times p$ matrix $\mathbf{M}$ whose element $m_{ij}$ is defined as $\sum_{k=1}^m a_{ik}b_{kj}$. %
If we represent matrices as relations with triples (row number, column number, value), then the \Rel definition of matrix multiplication mimics its mathematical definition: 
\begin{centeredrel*}
def MatrixMult[{A},{B},i,j] : sum[ [k] : A[i,k]*B[k,j] ]  
\end{centeredrel*}
Given two matrices \relverb+M1+ and \relverb+M2+, \relverb+MatrixMult[M1,M2]+ evaluates to the relation that represents their product. So, not only is this definition similar to the mathematical definition and easy to
program, it is also suitable as a library definition, since it takes
relations as parameters. %
It perfectly fits the paradigm of growing a language from a small core,
making it easy for users to expand that core %
using library functions. %
Furthermore, the relational model's data independence principle makes it possible for a \Rel engine to automatically choose the right data structures for \relinline{M1} and \relinline{M2} depending on whether they are dense or sparse, in-memory or not, etc.

As a second example, consider the following definition of all pairs shortest paths ({APSP}), given sets of nodes \relinline{V} and edges \relinline{E}:
\begin{centeredrel*}
def APSP({V},{E},x,y,0) : V(x) and V(y) and x = y
def APSP({V},{E},x,y,i) : 
i = min[{(j): exists((z) | E(x,z) and APSP(V,E,z,y,j-1))}]
\end{centeredrel*}
We can read the code as follows. The shortest path from $x$ to $y$ has length 0 if $x$ and $y$ are nodes and $x = y$. Otherwise, the shortest path length $i$ is the minimum $j$ such that an out-neighbor $z$ of $x$ has a shortest path of length $j-1$ to $y$. Again, \relverb+APSP+ can serve as a library definition in the sense that, if we have a directed graph with nodes \relverb+N+ and edges \relverb+NN+, and two nodes $u$ and $v$, we can call \relverb+APSP[N,NN,u,v]+ to give us the length of the shortest path from $u$ to $v$.

\subsection*{Rel is a Relational Programming Language}
Functional programming languages use functions as the main building block. Similarly, imperative programming languages use procedures. In \Rel, this role is fulfilled by relations. In principle, the relational approach subsumes the functional approach because every function is a relation. 

We illustrate this by showing how \Rel's notion of \emph{relational application} generalizes function application.
For simplicity, consider a function $f : V \times V \to V$, where $V$ is a domain of values. Such a function can be represented as a ternary relation $F$ consisting of the triples $(a,b,f(a,b))$ in which the first two columns determine the third. \Rel's syntax \relinline{F[a,b]} corresponds to the case where we provide $f$ with two parameters $a$ and $b$ and obtain $f(a,b)$ as a result. But \Rel also allows writing \relinline{F[a]} to return all pairs $(b,f(a,b))$ or \relinline{F[a,b,c]} to return true if and only if $(a,b,c) \in F$ (in other words, if and only if $c = f(a,b)$). Likewise, it is possible to write \relinline{F} to return the entire relation $F$.

\paragraph{Paper Overview}
In Section~\ref{sec:gnf}, we explain the principles of modeling in \GNFfull{} (GNF). Then, we move to the fundamental ingredients of \Rel in Section~\ref{sec:basics} and move to the features that give \Rel the capability to do programming in the large in Section~\ref{sec:programming-in-the-large}. In Section~\ref{sec:growing} we show how relational algebra, linear algebra, and graph algorithms can be implemented as libraries. We explain the ideas behind building relational knowledge graphs using \Rel in Section~\ref{sec:relational-knowledge-graph}.
Section~\ref{sec:rel:past-present-future} discusses \Rel influences, its use, and future plans. We present a formal semantics of (a  core of) \Rel in Addendum~\ref{sec:maintext-sem}.

\section{Data Modeling and Graph Normal Form}\label{sec:gnf}

Codd's original papers \cite{Codd70,Codd71,Codd79,Codd82} introduced a model based on $n$-ary relations, a normal form for database relations, and the concept of a universal sublanguage. The $n$-tuples in each relation represent facts about the domain being modeled.   
Since a fact can hold neither multiple times, nor partially, the pure relational model  
\cite{Codd70} has neither multiplicities (bags) nor nulls, i.e., it is based on set semantics and two-valued predicate logic. 
\Rel takes this one step further: tuples represent facts that are {\em  indivisible}. A complex fact like \emph{``Edgar was born on 19 August 1923 in Underhill''} is better thought of as two indivisible facts, \emph{``Edgar was born on 19 August 1923''} and \emph{``Edgar  was born in Underhill''}. 
\Rel shares this perspective with fact-based modeling \cite{orm-book}, inspired by the same work referenced by Codd~\cite{Codd79}. As facts involve real-world concepts rather than database constants (\emph{things, not strings}~\cite{things}), 
tuples should store unique, context-independent representations of these concepts.
In the above example,  \emph{Underhill} is the area on the Isle of Portland in Dorset (UK), not the town in Wisconsin nor the travel name of Frodo Baggins; the database should be able to distinguish between  them all. %
The combination of indivisibility of facts and the \emph{things, not strings} paradigm gives rise to \Rel's \emph{\GNFfull{} (\GNF)}, as we explain below.

\paragraph{Indivisibility of facts.}
Traditional modeling methodologies take a record-based perspective, in which a tuple may represent an entity with multiple attributes, such as a person with a date and place of birth. 
One can reconcile this with \Rel's fact-based perspective, and ensure the indivisibility of facts principle,  by assuming a higher degree of normalization, 
by enforcing \emph{sixth normal form (6NF)}~\cite{date-darwen-lorentzos}.
Indeed, GNF requires each relation to be in 6NF, which means that:
\begin{itemize}
\item the set of all its columns is its unique key, or
\item the set of all its columns except one is its unique key. 
\end{itemize}
We can view such a relation as either a set of distinct composite keys $\bar k$, or a set of key-value pairs $(\bar k, v)$ representing a function that maps keys $\bar k$ to atomic values $v$. (Indeed, in \Rel we assume that if there is a non-key column, it is the last one.)

For instance, consider a simple conceptual model shown below as an Entity-Relationship (ER) diagram.

 \smallskip
\begin{center}
\includegraphics[scale=0.54]{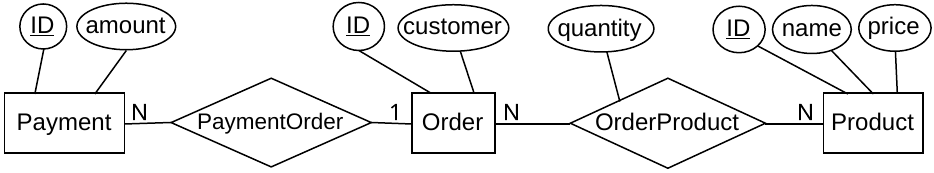}
\end{center}
 \smallskip

\noindent Orders may involve multiple products ordered in some quantities. Multiple payments can be made for each order. This conceptual schema leads to the following GNF database schema (key attributes are underlined):
{\small \texttt{
\begin{itemize}
    \item[] ProductPrice(\underline{product}\,,\,price)
    \item[] ProductName(\underline{product}\,,\,name)
    \item[] OrderCustomer(\underline{order}\,,\,customer)
    \item[] OrderProductQuantity(\underline{order}\,,\underline{product}\,,quantity)
    \item[] PaymentAmount(\underline{payment}\,,amount)
    \item[] PaymentOrder(\underline{payment}\,,order) 
\end{itemize}}}

\sloppy
\noindent Note that if we had a relation {\small \texttt{Product(\underline{product}\,,\,name,\,price)}}, it would not be in GNF, as neither {\small \texttt{name}} nor {\small \texttt{price}} are key attributes. This is why we have two relations: %
{\small \texttt{ProductName}} and {\small \texttt{ProductPrice}}, which store atomic facts about products.

\paragraph{Things, not strings.}
\GNF relies on a conceptual model that distinguishes between \emph{entities} (products, orders, etc.) and \emph{values} (integers, dates, etc.).
In relations, values are represented by themselves as usual, whereas entities are represented by internal \emph{identifiers}. In \GNF, identifiers are disjoint from values and every entity in the database is represented by an identifier that is unique within the entire database. 
So, \GNF does not allow disjoint concepts, such as product and order, to have the same identifier in the database. We call this the \emph{unique identifier property}.

\medskip

In summary, \emph{\GNFfull (\GNF)} comprises the following conditions:
\begin{enumerate}[(1)]
\item for each $k$-ary relation~$R$,
\begin{itemize}
\item all $k$ columns of~$R$ are the key, or
\item the first $k-1$ columns of~$R$ are the key for~$R$;
\end{itemize}
\item the unique identifier property holds.
\end{enumerate}
Condition (1) captures indivisibility of facts, and Condition (2) the \emph{things, not strings} paradigm.
\textcolor{black}{
Relational databases in \GNF{} can be thought of as \emph{Relational Knowledge Graphs}. We  explain differences between them and other models of knowledge graphs in Section~\ref{sec:relational-knowledge-graph}.
}

Under GNF, each  relation is a set. Moreover, there is no need for nulls: rather than using a null in the non-key column, we simply omit the whole tuple (possibly using a separate relation for keys alone). 
Other benefits of GNF such as semantic stability and support for temporal features are discussed in~\cite{ArefCGKOPVW15}. This level of normalization also allows dropping column names,
because relation names alone (under the naming scheme we use) are sufficiently informative; for instance, compare {\small\texttt{Product(product,name,price)}} with {\small\texttt{ProductName}} and {\small\texttt{ProductPrice}}.

\newcolumntype{C}{>{\centering\arraybackslash}X}

\newcommand{\ourdb}{
\begin{figure*}[t]
\begin{center}
  \begin{tabular}[t]{cc}
    \multicolumn{2}{c}{\textit{PaymentOrder}}\\
    \toprule
    "Pmt1"& "O1" \\
    "Pmt2"& "O2" \\
    "Pmt3"& "O1" \\
    "Pmt4"& "O3" \\
  \end{tabular}
  \quad
  \quad
  \quad
  \begin{tabularx}{26mm}[t]{%
    C>{\centering\arraybackslash}p{7mm}%
    }
    \multicolumn{2}{c}{\textit{PaymentAmount}}\\
    \toprule
    "Pmt1"& 20 \\
    "Pmt2"& 10 \\
    "Pmt3"& 10 \\
    "Pmt4"& 90 \\
  \end{tabularx}
  \quad
  \quad
  \quad
  \begin{tabularx}{35mm}[t]{%
  >{\centering\arraybackslash}p{10mm}%
  C%
  >{\centering\arraybackslash}p{6mm}}
    \multicolumn{3}{c}{\textit{OrderProductQuantity}}\\
    \toprule
     "O1" & "P1" & 2 \\
     "O1" & "P2" & 1 \\
     "O2" & "P1" & 1\\
     "O3" & "P3" & 4\\
  \end{tabularx}
  \quad\quad\quad
  \begin{tabularx}{20mm}[t]{CC}
    \multicolumn{2}{c}{\textit{ProductPrice}}\\
    \toprule
     "P1" & 10 \\
     "P2" & 20 \\
     "P3" & 30 \\
     "P4" & 40 \\
  \end{tabularx}

\end{center}
\caption{Example database containing orders, products included in orders (with their amount), and product prices.\label{fig:database}}
\end{figure*}
}

\ourdb
\section{Basics}
\label{sec:basics}
To introduce basic features of \Rel, we use the database in Figure~\ref{fig:database}, which comprises a subset of relations from the example in Section~\ref{sec:gnf}.
We will use the identifiers %
$\text{"Pmt1"},\text{"Pmt2"}, \dots$ for payments, $\text{"P1"},\text{"P2"},\dots$ for products, and
$\text{"O1"},\text{"O2"},\dots$ for orders.
These need to be disjoint to meet the {\em unique identifier property} (condition (2) of \GNF), which is illustrated by the identifier naming scheme.
Recall that the \Rel data model does not associate names with the attributes of relations, and refers to them by their position only. As explained in Section~\ref{sec:gnf}, under GNF, SQL columns essentially give rise to separate relations, which means that SQL column names are naturally represented in relation names in \Rel (under the adopted naming scheme).

\subsection{Datalog as a Starting Point}
\label{sec:expressions}
The starting point of \Rel is Datalog rules with first-order formulas in their bodies. 
We assume a very basic level of familiarity with the relational data model (relational algebra and calculus)~\cite{cowbook,ABLMP21}. 
A \Rel program is a set of rules, the most basic of which has the form 
\begin{centeredrel}[label=eq:datalog-rule]
def RName( $\sexpr{VariableList}$ ) : $\sexpr{RelExpression}$
\end{centeredrel}
The rule is structured like a Datalog rule: \relinline{RName} is a relation name, $\sexpr{VariableList}$ is a list of variables, and $\sexpr{RelExpression}$ is an expression that evaluates to a result that gives meaning to the variables in $\sexpr{VariableList}$.
For example, consider the rule 
\begin{centeredrel*}
def OrderWithPayment(y) : exists ((x) | PaymentOrder(x,y))
\end{centeredrel*} 
\sloppy
which adds to \textit{OrderWithPayment} the tuples 
$\langle \text{"O1"}  \rangle, \langle \text{"O2"} \rangle,\langle \text{"O3"} \rangle$, 
that is, 
the orders that have received at least one payment. \textcolor{black}{Throughout, we write tuples using angular brackets (for example, $\langle \text{"O1"}  \rangle$ is a unary tuple, and $\langle 1,2,3\rangle$ is a ternary tuple), and  fonts such as \relinline{R} for variables and relation names in \Rel, while italics are used to refer to an extent $\textit{R}$ of \relinline{R}.} %
\Rel uses \textbf{set semantics}: in the above example, 
"O1" only occurs once in the result, even though this order received two payments.

Notice that the rule uses an existentially quantified variable \relinline{x} that is not used elsewhere. 
\Rel allows anonymous variables, denoted ``\relinline{_}'', to simplify the syntax. It is equivalent to write
\begin{centeredrel*}[label=rule:PaymentDatesWildcard]
def OrderWithPayment(y) : PaymentOrder(_,y)
\end{centeredrel*} 
We note that different occurrences of \relinline{_} can bind to different values. For example,
\begin{centeredrel*} 
def OrderedProducts(y) : OrderProductQuantity(_,y,_)
\end{centeredrel*}
computes the products that were ordered. In particular, we get $\langle \text{"P1"}  \rangle, \langle \text{"P2"} \rangle,\langle \text{"P3"} \rangle$ as the result.

As usual, repeated variables express join conditions: the rule
\begin{centeredrel*}
def OrderedProductPrice(x,y) : 
           OrderProductQuantity(_,x,_) and ProductPrice(x,y)
\end{centeredrel*} 
takes the ordered products and adds them with their price into \textit{OrderedProductPrice}. This is the set $\{\langle \text{"P1"}, 10 \rangle$, $\langle \text{"P2"}, 20 \rangle$, $\langle \text{"P3"}, 30 \rangle\}$. 

Finally, $\sexpr{RelExpression}$ allows both existential and universal quantification, as well as all Boolean operations.
The rule 
\begin{centeredrel*}
def NotOrdered(x) : ProductPrice(x,_) and 
    not exists ((y1,y2) | OrderProductQuantity(y1,x,y2))
\end{centeredrel*} 
adds the products that were not ordered to \textit{NotOrdered}. The rule
\begin{centeredrel*}
def NotOrdered(x) : ProductPrice(x,_) and 
    forall ((y1,y2) | not OrderProductQuantity(y1,x,y2))
\end{centeredrel*} 
is equivalent: both add \text{"P4"} to \textit{NotOrdered}. 

Using wildcards, we can equivalently define \textit{NotOrdered} as 
\begin{centeredrel*}
def NotOrdered(x) : 
    ProductPrice(x,_) and not OrderProductQuantity(_,x,_)
\end{centeredrel*} 
This means that \relinline{_} is equivalent to an anonymous variable that is existentially quantified immediately outside of the atom where it is used.
It is possible to restrict the range of quantifiers. For example, if we have a (database or pre-computed) relation $V = \{\text{"O1}, \text{"O2"}\}$, we can write
\begin{centeredrel*}
def AlwaysOrdered(x) : ProductPrice(x,_) and
    forall ((o in V) | OrderProductQuantity(o,x,_))
\end{centeredrel*}
to obtain the set of products that were in every order in $V$.
\Rel allows Boolean connectives  \relinline{implies}, \relinline{iff}, and \relinline{xor} as syntactic sugar, with their usual meanings. 

\paragraph{Safety} Notice that negation can lead to  \emph{safety} issues. These occur, for example, when we cannot limit the number of results to a query such as 
\begin{centeredrel*}
  def NotP1Price(x) : not ProductPrice("P1",x)
\end{centeredrel*}
It computes the set of prices that product "P1" does not have, which is (conceptually) infinite.
{Relational  calculus and SQL use \emph{range restriction} to circumvent this kind of problem. Specifically, variables can only range over elements in the database or those constructed from database entries by means of expressions, thereby rendering ranges of variables finite. 
\Rel takes a more flexible approach, allowing unsafe subexpressions as long as the whole expression is safe, and reasons about safety of expressions using a set of rules, based on  \cite{GuagliardoLMMMPS25}. Since safety is an undecidable condition \cite{ABLMP21}, it is impossible  to distinguish all safe and unsafe expressions. \Rel takes a conservative approach, ensuring that the engine never attempts to evaluate an expression that could be unsafe.

\subsection{Infinite Relations}

\Rel makes it possible to use (conceptually) infinite relations. For example, \relinline{Int(x)} tests if \relinline{x} is an integer. %
Another example is the ternary relation \relinline{add}, which contains all triples $\langle x, y ,z \rangle$ of all data types such that $x + y = z$. Using these, one can write rules such as
\begin{centeredrel*}
def DiscountedProductPrice(x,y) : 
    exists ((z) | ProductPrice(x,z) and add(y,5,z))
\end{centeredrel*}
which computes the $DiscountedProductPrice$ relation in which every product received a discount of 5, that is, $\{\langle \text{"P1"},5 \rangle, \allowbreak \langle \text{"P2"}, \allowbreak 15 \rangle, \allowbreak \langle \text{"P3"}, 25 \rangle , \langle \text{"P4"}, 35 \rangle\}$.
Again, using such infinite relations requires some care, because queries may be unsafe, i.e., return conceptually infinite results. One such unsafe example is
\begin{centeredrel*}
def AdditiveInverse(x,y) : Int(x) and Int(y) and add(x,y,0)
\end{centeredrel*}
Indeed, it asks for the pairs of integers $(x,y)$ whose sum is zero, which is an infinite set. \Rel's aforementioned set of safety rules~\cite{GuagliardoLMMMPS25}
will detect that this expression is potentially infinite. Still such expressions can be written and used in other queries; for example, an expression that intersects \relinline{AdditiveInverse} with a finite set will be seen as safe and thus evaluated to produce a finite result.

\paragraph{Arithmetic}

\Rel supports all standard arithmetic operators such as addition, multiplication, division, modulo, etc., using the standard infix notation. For example,  one can write 
\begin{centeredrel*}
def PsychologicallyPriced(x) :
    exists ((y) | ProductPrice(x,y) and y % 100 = 99)
\end{centeredrel*}
to find products whose prices are 99 modulo 100. 
Each arithmetic operator has an equivalent ``relational'' notation. For instance, \relinline{add} is the relational notation for \relinline{+}, \relinline{multiply} is the relational notation for \relinline{*}, and \relinline{modulo} is the relational notation for $\%$.

\subsection{Code Flow and Recursion}
\label{sec:flow}

Analogously to Datalog programs, rules can be written in any order. The ordering of rules in \Rel programs has no effect on their semantics. The following small program computes products that are ordered together with some expensive product. The rules are easiest to understand from top to bottom, but the program would compute the same result if the rules were ordered differently. 
\begin{centeredrel*}
def SameOrder(p1, p2) :
    exists((order) | OrderProductQuantity(order, p1, _)
                 and OrderProductQuantity(order, p2, _))
def SameOrderDiffProduct(p1, p2) :
    SameOrder(p1, p2) and p1 != p2
def Expensive(p) : 
    exists ((price) | ProductPrice(p,price) and price > 15)
def BoughtWithExpensiveProduct(p) : 
    exists((x in Expensive) | SameOrderDiffProduct(x, p))
\end{centeredrel*}
Here, \relinline{SameOrder} evaluates to the set of pairs of products bought together in the same order; \relinline{SameOrderDiffProduct} limits those to pairs of distinct products and evaluates to $\{\langle \text{"P1"},\text{"P2"}\rangle, \langle \text{"P2"},\text{"P1"}\rangle\}$. Then, \relinline{Expensive} evaluates to all products whose price is more than 15, and \relinline{BoughtWithExpensiveProduct} evaluates to the set of products that were bought together with an expensive product ("P1").

\paragraph{Recursion} Being squarely rooted in Datalog, \Rel allows recursion. For example, assume that we have a binary relation $\textit{E}$ of edges in some graph. Then the program
\begin{centeredrel*}
def TC_E(x,y) : E(x,y) 
def TC_E(x,y) : exists((z) | E(x,z) and TC_E(z,y)) 
\end{centeredrel*}
computes the transitive closure of \relinline{E}, that is, the node pairs $x,y$ such that $y$ is reachable from $x$ using the edges in \textit{E}. We note that recursion in \Rel does not need to be linear, that is, \relinline{TC_E} is allowed to occur multiple times on the same right-hand side of a rule.

\paragraph{Rules Defining the Same Relation Name} 
 \Rel allows multiple rules with the same relation name on the left-hand side, such as our example for \relinline{TC_E}. The semantics of multiple such rules is similar to Datalog. Indeed, having two rules such as
\begin{centeredrel*}
def ID ( $\sexpr{VariableList}$ ) : $\sexpr{RelExpression1}$
def ID ( $\sexpr{VariableList}$ ) : $\sexpr{RelExpression2}$
\end{centeredrel*}
is equivalent to 
\begin{centeredrel*}
def ID ( $\sexpr{VariableList}$ ) : $\sexpr{RelExpression1}$ or $\sexpr{RelExpression2}$
\end{centeredrel*}
that is, the union of the results of the two rules.

\paragraph{Giving Meaning to Recursive Rules} Although the program for transitive closure is fairly easy to understand, this is much less so for general recursive programs that involve negation. In general, the semantics is defined based on a \emph{dependency graph} of the program, which is divided into so-called \emph{strata} by non-monotonic operators, such as negation. The semantics of \Rel 
is consistent with the \emph{stratified semantics of Datalog} \cite[Chapter 15]{ahv_book}, but \Rel also allows non-stratified programs, {see Addendum~\ref{sec:maintext-sem}}.

\subsection{Output and Updates}

A \Rel \emph{query} (or program) is a sequence of \Rel rules defining some relations, typically referring to \emph{base relations} stored in the database (e.g., \relinline{ProductPrice}, \relinline{OrderProductQuantity}, \relinline{PaymentOrder}, and \relinline{PaymentAmount}). The execution of a query against a database is called a \emph{transaction}. 
A transaction performs 
computation using \emph{derived relations} and interacts with the environment using \emph{control relations}. The latter are just reserved relation names, such as \relinline{insert}, \relinline{delete}, and \relinline{output}. They are defined using rules and computed just like any other relation, but they have a special purpose. 

The control relation \relinline{output} specifies query answers: when a query is issued, what gets returned is the computed content of the relation \relinline{output}.   For instance, the  query
\begin{centeredrel*}
def output(x) : exists( (y) | ProductPrice(x,y) and y > 30) 
\end{centeredrel*}outputs all products whose price exceeds 30. Note that \relinline{output} is defined as any other relation; its side effect is that its contents are displayed to the user.

A transaction can modify the content of the database using control relations \relinline{insert} and \relinline{delete}. 
Assume we have binary relations \relinline{OrderTotal} and \relinline{OrderPaid}  containing total prices and total payments of orders (these will be defined with the help of aggregation in Section \ref{sec:aggregation}). The following  deletes information about fully paid orders by adding the rule
\begin{centeredrel*}
def delete(:OrderProductQuantity,x,y,z) : 
      OrderProductQuantity(x,y,z) and 
      exists( (u) | OrderPaid(x,u) and OrderTotal(x,u) )
\end{centeredrel*}
and 
inserts the fully paid orders to the base relation \relinline{ClosedOrders} by adding the rule
\begin{centeredrel*}
def insert(:ClosedOrders,x) : 
    exists( (u) | OrderPaid(x,u) and OrderTotal(x,u))
\end{centeredrel*}
(here the use of \relinline{:} in front of a relation name indicates passing the name of a relation as a parameter).
There is no need to  declare a new base relation: if \relinline{ClosedOrders} does not exist, it will be created on the spot.

When a transaction terminates, changes to the database are persisted, unless the transaction is aborted (for instance, when integrity constraints are violated; more on this in Section~\ref{sec:constraints}).

\subsection{Integrity Constraints}\label{sec:constraints}

Integrity constraints check whether  relations comply with specified requirements. %
They are specified using  \relinline{ic} and \relinline{requires} keywords. Common types of constraints are type constraints which are used to ensure that values conform to a type. For instance, to ensure that quantities of ordered products are integers we write 
\begin{centeredrel*}
ic integer_quantities() requires 
    forall((x) | OrderProductQuantity(_,_,x) implies Int(x))
\end{centeredrel*}
where \relinline{Int} is the predicate that tests if its argument is an integer %
To retrieve the entries that violate this constraint, we add a parameter: 
\begin{centeredrel*}
ic integer_quantities(x) requires 
    OrderProductQuantity(_,_,x) implies Int(x)
\end{centeredrel*}
Then \relinline{integer_quantities} will be populated with the values $x$ that violate the constraint.

Integrity constraints can be used to express functional dependencies, foreign keys, etc.
The following verifies that each product in the \relinline{OrderProductQuantity} appears in the \relinline{ProductPrice} relation:
\begin{centeredrel*}
ic valid_products(x) requires 
    OrderProductQuantity(_,x,_) implies ProductPrice(x,_)
\end{centeredrel*}
If a transaction violates a constraint, it is aborted. 

\section{Enabling Programming in the Large}\label{sec:programming-in-the-large}

We now discuss language features that, together with the basics from Section~\ref{sec:basics}, enable \Rel to do programming in the large, giving it the power to define libraries. For this, we need mechanisms for passing parameters that are more complex than individual values.

\subsection{Tuple Variables}
\label{sec:varargs}
\label{sec:tuple-vars}
Assume that we have binary relations $R = \{\langle 1,2 \rangle, \langle 3,4 \rangle\}$ 
and  $S = \{\langle 5, 6 \rangle\}$, and we want to compute their Cartesian product, defined as the set of tuples $\langle a,b,c,d \rangle$ such that $\langle a,b \rangle \in R$ and $\langle c,d \rangle \in S$. We could write this as
\begin{centeredrel*}
    def ProductRS(a,b,c,d) : R(a,b) and S(c,d)
\end{centeredrel*}
This works fine and will add two tuples to \textit{ProductRS}. But what if $R$ were ternary and $S$ binary? We would have to write
\begin{centeredrel*}
    def ProductRS(a,b,c,d,e) : R(a,b,c) and S(d,e)
\end{centeredrel*}

Writing such code for all different arities is not only repetitive and error-prone, but also incomplete, because there is a maximal arity for which the code works. This is why \Rel allows \emph{tuple variables}, which are syntactically distinguished from ordinary variables by trailing dots.
Using %
them, we can compute the Cartesian product of \relinline{R} and \relinline{S} without knowing  the arities of the tuples that are in them:
\begin{centeredrel*}
    def ProductRS(x...,y...) : R(x...) and S(x...)
\end{centeredrel*}
Here, the variables \relinline{x...} and \relinline{y...} can bind to arbitrary-length parts of tuples (including length zero).

Using tuple variables, one can produce relations of tuples of different arities; for example
\begin{centeredrel*}
    def Prefix(x...) : R(x...,_...)
\end{centeredrel*}
computes all prefixes of the tuples in \relinline{R}. Here, \relinline{_...} plays the role of  a wildcard for tuples: it matches an arbitrary tuple of arbitrary arity, again including arity zero. 

Tuple variables are a powerful tool, especially if combined with recursion; for example,
\begin{centeredrel*}
def Perm(x...) : R(x...)
def Perm(x...,a,y...,b,z...) : Perm(x...,b,y...,a,z...)
\end{centeredrel*}
computes all permutations of tuples in \relinline{R}, using the fact that each permutation is a product of transpositions.

\subsection{Relation Variables}
\label{sec:higher-order}

Being able to define \relinline{ProductRS} independently of the arities of \relinline{R} and \relinline{S} is nice, but it would be even better if we could pass \relinline{R} and \relinline{S} as parameters to a more general operator that computes the Cartesian product. For example, we would like \relinline{Product[R,S]} to return the Cartesian product of \relinline{R} and \relinline{S}. We will achieve this in two steps: the first is \emph{relation variables}, the second is \emph{relational application}.

In \Rel syntax, relation variables in the head of rules are indicated by enclosing them in curly brackets:
\begin{centeredrel*}
def Product({A},{B},x...,y...) : A(x...) and B(y...)
\end{centeredrel*}
Using this definition, we can write \relinline{Product(R,S,a,b,c,d)} to test if $\langle a,b,c,d \rangle$ is in the Cartesian product of \relinline{R} and \relinline{S}.

Until now, every rule of the form \eqref{eq:datalog-rule} defined a \emph{relation}, that is, a set of tuples. But what about \relinline{Product}? In fact, this is also a relation, but it is conceptually second-order. Its first two columns contain (first-order) relations instead of values:

\begin{center}
\begin{tabular}{cccccc}
\multicolumn{6}{c}{\it Product}\\
\toprule
$\{0\}$ & $\{0\}$ & 0 & 0 & & \\
$\{0\}$ & $\{1\}$ & 0 & 1 \\
$\{1\}$ & $\{0\}$ & 1 & 0\\
$\cdots$\\
$\{\langle 1,2\rangle, \langle 3, 4 \rangle\}$ & $\{\langle 5, 6 \rangle\}$ & 1 & 2 & 5 & 6\\
$\{\langle 1,2\rangle, \langle 3, 4 \rangle\}$ & $\{\langle 5, 6 \rangle\}$ & 3 & 4 & 5 & 6\\
$\dots$
\end{tabular}
\end{center}

\noindent Notice that \textit{Product} is infinite. It has infinitely many rows, because it contains infinitely many relations in its first column alone. It also has infinitely many columns, because the relations in the first two columns do not have an upper bound on their arity. As such, the tuples in their Cartesian product become arbitrarily long.

\subsection{Relational Application}\label{sec:relational-application}

We can already write \relinline{OrderProductQuantity("O1","P1",2)} to test if $\langle \text{"O1"}, \text{"P1"}, 2 \rangle \in \textit{OrderProductQuantity}$. This is the standard notation in our field for \emph{relational atoms}~\cite{ABLMP21,cowbook}.
Another, more general way of understanding this is seeing \relinline{OrderProductQuantity} as a Boolean function that, given three input parameters, $a$, $b$, $c$, returns whether $\langle a,b,c \rangle \in \textit{OrderProductQuantity}$. This principle works for all kinds of relations that we have seen, including second-order relations. For example, using $R$ and $S$ from Section~\ref{sec:tuple-vars},
\begin{centeredrel*}
    Product(R, S, 1, 2, 5, 6)
\end{centeredrel*}
evaluates to true, since the second-order tuple $\langle R, \allowbreak S, \allowbreak 1, 2, 5, 6 \rangle$ is in the relation \textit{Product}.
We just described one form of \emph{relational application} in \Rel. 
We call it \emph{full (relational) application}, because the syntax with \relinline{( )} requires that all arguments to the Boolean function are produced in order for it to evaluate. 

\Rel also supports \emph{partial (relational) application}, which uses \relinline{[ ]} instead of \relinline{( )}. Partial application returns all suffixes in a relation that are consistent with a given prefix. For example, \relinline{OrderProductQuantity["O1"]} evaluates to $\{\langle \text{"P1"}, 2 \rangle, \langle \text{"P2"}, 1 \rangle\}$, because $\langle \text{"O1"},\text{"P1"},2 \rangle$ and $\langle \text{"O1"},\text{"P2"},1 \rangle$ are the tuples in \textit{OrderProductQuantity} that start with "O1". 

The same principle holds for second-order relations. Hence, 
\begin{centeredrel*}
Product[R, S]
\end{centeredrel*}
evaluates to the Cartesian product of the relations $R$ and $S$. In fact, \Rel has a special notation for this ubiquitous operation: \relinline{(R,S)}. For example, \relinline{(PaymentOrder,ProductPrice)} is the Cartesian product of \relinline{PaymentOrder} and \relinline{ProductPrice}, while \relinline{("P4",40)} is the relation containing a single tuple $\langle$\relinline{"P4", 40}$\rangle$. 

\textcolor{black}{Partial application does not need to provide all arguments, and thus evaluates to a relation. If this relation has arity zero, the result is either $\{\langle \rangle\}$ (the relation with the empty tuple) or $\{\}$ (the empty relation). In \Rel, these encode Boolean values true and false: true is encoded as $\{\langle \rangle\}$   and false as $\{\}$, respectively (as usual in the relational data model~\cite{ABLMP21}). Therefore, partial application is identical to full application if all arguments are provided.}

\subsection{Abstraction}

Recall, \Rel rules are of the form \eqref{eq:datalog-rule}: 
\relinline{def RName(}{\small$\sexpr{VariableList}$}\relinline{)}\relinline{:}{\small$\sexpr{RelExpression}$}.
Actually, \Rel rules have a more general form, which is 
\begin{centeredrel}[label=eq:rule]
def RName $\sexpr{Abstraction}$
\end{centeredrel}
where $\sexpr{Abstraction}$ can have one of the following forms:
\begin{subequations}
\begin{centeredrel}[label=eq:abstraction-round]
    { ( $\sexpr{VariableList}$ ) : $\sexpr{RelExpression}$ }
\end{centeredrel}
\vspace{-1ex}
\begin{centeredrel}[label=eq:abstraction-square]
    { [ $\sexpr{VariableList}$ ] : $\sexpr{RelExpression}$ }
\end{centeredrel}
\end{subequations}
and the outer curly braces can be omitted to allow the form \eqref{eq:datalog-rule}.
In the general form \eqref{eq:rule}, the result of $\sexpr{Abstraction}$ is evaluated and added to \textit{RName}. We say \emph{added to}, because there can be multiple rules with the same \relinline{RName} on their left hand side. We now explain how $\sexpr{Abstraction}$ works.

In abstractions of the form~\eqref{eq:abstraction-round}, the $\sexpr{RelExpression}$ on the right hand side should evaluate to a Boolean. In fact, \Rel only allows a syntactic subset of $\sexpr{RelExpression}$ on the right hand side, namely $\sexpr{Formula}$, which guarantees evaluation to a Boolean. The form~\eqref{eq:abstraction-square} allows general expressions instead of Boolean conditions right of the colon.

Abstractions of the form \eqref{eq:abstraction-round} are inspired by \emph{set comprehension} from mathematics where we write, e.g., $\{n \in \nat : \exists m \in \nat$ such that $n = 2m\}$ for the set of even numbers. For example, 
\begin{centeredrel*}
{(x,y) : OrderProductQuantity(x,"P1",y)}
\end{centeredrel*}
evaluates to the set of orders and quantities of orders of product \text{"P1"}. Set comprehensions can be used to perform selection and projection, see Section~\ref{sec:RA}.

The difference with \eqref{eq:abstraction-square} is that we now use square brackets left of the colon. For example,\footnote{\Rel uses the comma as an infix operator to denote the Cartesian product, see Sections~\ref{sec:relational-application} and \ref{sec:RA}.} 
\begin{centeredrel}[label=eq:abstraction-example]
{[x,y] : (OrderProductQuantity[x], PaymentOrder(y,x))}
\end{centeredrel}
works as follows. For each possible pair $\langle v_x, v_y \rangle$ of values for \relinline{x} and \relinline{y}, we evaluate the expression \relinline{(OrderProductQuantity[x]}, \relinline{PaymentOrder(y,x))} which gives products and their quantities in the order $v_x$ for which a payment $v_y$ was made. Whenever the result of the evaluation of this expression is a nonempty set 
 $S$ of tuples, we include $\{\langle v_x, v_y \rangle\} \times S$ in the answer. 
 For example, for values $v_x=\text{"O1"}$ and $v_y=\text{"Pmt1"}$, this set $S$ contains the result of \relinline{OrderProductQuantity["O1"]}, i.e., $\{\langle \text{"P1"},2\rangle, \langle \text{"P2"}, 1\rangle\}$.
 Thus, tuples 
 $\langle \text{"O1"}, \text{"Pmt1}, \text{"P1"},2\rangle$ and $\langle \text{"O1"}, \text{"Pmt1}, \text{"P2"}, 1\rangle$ will be included in the result.  Similar to quantification, we can restrict the range of variables on the left hand side to a finite set. If \relinline{V}$ = \{\langle \text{"Pmt2"} \rangle, \langle \text{"Pmt4"} \rangle\}$,
\begin{centeredrel*}
{[x, y in V] : (OrderProductQuantity[x], PaymentOrder(y,x))}
\end{centeredrel*}
returns only results of the previous query pertaining to payments "Pmt2" and "Pmt4"; i.e., $\langle \text{"O2"}, \text{"Pmt2}, \text{"P1"}, 1 \rangle$ and $\langle \text{"O3"}, \text{"Pmt4"}, \text{"P3"} ,\text{4} \rangle$.

We call this operation \emph{(relational) abstraction} to draw a parallel with functional abstraction in functional programming languages. Indeed, lambda-abstraction $\lambda x. e$ denotes a function that for argument $x$ computes $e(x)$. This function can be viewed as a relation, namely the set of pairs $(x,e(x))$. In relational abstraction, the difference is that $e$ can be an expression that produces a relation instead of a function. The semantics of the abstraction then is the set of tuples $(x,y_1,\ldots,y_k)$ with $(y_1,\ldots,y_k) \in e(x)$.

\section{Growing the Language}\label{sec:growing}
In this section we discuss how the language constructs in Section~\ref{sec:programming-in-the-large} enable \Rel to grow. We discuss different libraries of \Rel, starting with its standard library, and then explaining how to implement relational algebra, linear algebra, and graph analytics operations. The code examples will also illustrate several prominent features of \Rel programming. 
In Section \ref{sec:aggregation} we show 
how \Rel does aggregation under  set semantics, and explain that bag semantics is not necessary for correct computation of aggregate functions. In Section \ref{sec:linear-algebra} we  explain how to guard variables by limiting them to a finite domain, and in Section \ref{sec:graphlib} we show how to recurse until a stop condition is met. These examples illustrate how features that, under the sublanguage paradigm (e.g., in SQL), would require language extensions can simply be defined as libraries.

\subsection{Standard Library}

\Rel's standard library provides definitions of dozens of commonly used relations, ranging from trigonometric functions, through type and format conversions, to regex matching. Some of these relations are directly implemented in \Rel, like dot-join 
\begin{centeredrel*}
def dot_join({A},{B},x...,y...) : 
    exists((t) | A(x...,t) and B(t,y...))
\end{centeredrel*} 
which makes the join on the last position of \relinline{A} and the first position of \relinline{B} (dropping the join position in the result). Another example is left-override
\begin{centeredrel*}
def left_override({A},{B},x...) : A(x...)
def left_override({A},{B},x...,v) : 
    B(x...,v) and not A(x...,_)
\end{centeredrel*} 
which adds to \relinline{A} all tuples from \relinline{B} whose prefix obtained by dropping the last position 
does not appear as a prefix of a tuple in  \relinline{A}.

Others are just wrappers for external implementations, such as 
\begin{centeredrel*}
def log[x, y] : rel_primitive_log[x, y]
\end{centeredrel*}
These could be treated as language primitives, but in \Rel we prefer to think about them as library functions. Note that \relinline{add} and \relinline{multiply}, mentioned in Section~\ref{sec:basics} are also library relations defined likewise  using primitives \relinline{rel_primitive_add} and \relinline{rel_primitive_multiply}.

Recall that  \relinline{add} and \relinline{multiply} have corresponding infix operators. These are defined as follows in the library:
\begin{centeredrel*}
def (+)(x,y,z) : add(x,y,z)
def (*)(x,y,z) : multiply(x,y,z)
\end{centeredrel*} 

Other library relations have infix versions as well; for instance, we can use  \relinline{.} for \relinline{dot_join} and \relinline{<++} for \relinline{left_override}.
\subsection{Aggregation and Reduce}
\label{sec:aggregation}

In \Rel, aggregation is implemented as part of the standard library relying on a single additional primitive, exposed as a second-order  relation \relverb+reduce+. It is a ternary relation whose tuples are of the form $\langle F,R,v\rangle$ where $F$ is a binary operator, $R$ is a non-empty relation, and $v$ is a value. The meaning of \relinline{reduce(F,R,v)} is that $v$ is the value obtained by ``aggregating'' the values in the last column of $R$ using the operation $F$ (in the same way the {\small \texttt{reduce}} or {\small \texttt{fold}} operation works in many languages). 
The operations are performed in an arbitrary order, which means that~$F$ should be associative and commutative. Otherwise, the result could change from execution to execution. For example, relational aggregates can be defined as 
\begin{centeredrel*}
def sum[{A}] : reduce[add,A] 
def count[{A}] : reduce[add,(A,1)] 
def min[{A}] : reduce[minimum,A]
def max[{A}] : reduce[maximum,A]
def avg[{A}] : sum[A] / count[A]
\end{centeredrel*}
where \relverb{minimum[x,y]} and \relverb{maximum[x,y]} return minimum, resp. maximum of two numbers. 
 Note that \relverb+(A,1)+ includes $1$ at the end of each tuple in $A$; summing these ones gives the cardinality of \textit{A}.

Combining relational operators with aggregation we can define further operations such as, for example,  \relverb{Argmin} that on a set $A = \{(\bar{a}_1,v_1), \ldots, (\bar{a}_n,v_n)\}$ returns tuples $\bar{a}_i$ such that $v_i = \min_j v_j$:
\begin{centeredrel*}
def Argmin[{A}] : A.{min[A]}
\end{centeredrel*}

Aggregation is naturally combined with grouping. Suppose we want to sum up all payments for each order. We can then write:
\begin{centeredrel*}
def Ord(x) : OrderProductQuantity(x,_,_)
def OrderPaymentAmount(x,y,z) :      
     PaymentOrder(y,x) and PaymentAmount(y,z)
def OrderPaid[x in Ord] : sum[OrderPaymentAmount[x]] 
\end{centeredrel*}
Note, however, that orders without payments will not be included in the result, as for them \relinline{OrderPaymentAmount[x]} will evaluate to the empty set and, consequently,  so will \relinline{sum[OrderPaymentAmount[x]]}. To include them, one can use left override and write instead 
\begin{centeredrel*}
def OrderPaid[x in Ord] : sum[OrderPaymentAmount[x]] <++ 0
\end{centeredrel*}
which replaces the empty set by 0. 

\subsection{Relational and Linear Algebra: Point-Free Extensions}

Notations for linear algebra (LA) and relational algebra (RA) as well as languages such as APL~\cite{APL} and FP~\cite{FP} allow writing programs using a set of generally useful primitives and avoiding named variables, a style %
called point-free or tacit programming. The point-free style is also quite popular in business intelligence tools. Rel supports point-free style via libraries rather than language extensions, which we now demonstrate with libraries for RA and LA.

\subsubsection{Relational Algebra}\label{sec:RA}

A simple example of relations defined in the library are familiar relational algebra operators. Cartesian product is defined as the \relinline{Product} relation that we already discussed in Sections~\ref{sec:higher-order} and \ref{sec:relational-application}. Furthermore, as we already mentioned in Section~\ref{sec:higher-order}, \Rel uses the infix notation \relinline{(A,B)} to denote the Cartesian product of \relinline{A} and \relinline{B}.

The union of two relations can also be defined in the library:
\begin{centeredrel*}
def Union({A},{B},x...) : A(x...) or B(x...)
\end{centeredrel*}
Similarly to product, union has a special shorthand for it: \relinline{\{A; B\}}. 
This way we can build arbitrary relations from constants; e.g.,
\begin{centeredrel*}
    {(1,2,3) ; (4,5,6) ; (7,8,9)}
\end{centeredrel*}
evaluates to $\{\langle 1,2,3 \rangle, \langle 4,5,6 \rangle, \langle 7,8,9 \rangle\}$.
The remaining set operator of relational algebra --- difference --- is similarly defined:
\begin{centeredrel*}
def Minus({A},{B},x...) : A(x...) and not B(x...)
\end{centeredrel*}

A selection operator simply takes a relation and a condition --- which could be an infinite set --- and returns their intersection:
\begin{centeredrel*}
def Select({A},{Cond},x...) : A(x...) and Cond(x...)
\end{centeredrel*}

Consider, for example, a relational algebra expression $\sigma_{A_1=A_2}(R \times S) \cup B$, where relations $R$ and $S$ have attributes $A_1$ and $A_2$, respectively.
To express this in \Rel in point-free style, we first need an  infinite set of tuples whose first and second components are equal:
\begin{centeredrel*}
def Cond12(x1,x2,x...) : {x1=x2}
\end{centeredrel*}
and then define the entire expression as 
\begin{centeredrel*}
Union[Select[Product[R,S],Cond12],B]  
\end{centeredrel*}

\textcolor{black}{
Projection can be easily expressed in Rel using \emph{abstraction}.
For instance, the projection of a relation~$R$ on the first and third attribute may be expressed as:}
\begin{centeredrel*}
    (x,y) : R(x,_,y,_...)
\end{centeredrel*}

\paragraph{Expressions vs Formulas} %
We already know that some expressions always evaluate to Boolean values, which are $\{\langle\rangle\}$ (true) and $\{\}$ (false), the only two sets of tuples of arity zero. In \Rel, \sexpr{Formula} is a subclass of \sexpr{RelExpression} for which we can statically infer that only Boolean values are produced. These expressions can be combined with \relinline{and}, \relinline{or}, and \relinline{not}. Notice that for formulas, \relinline{and} is equivalent to the Cartesian product, and \relinline{or} to union: $F_1$ \relinline{and} $F_2$ is the same as \relinline{(}$F_1$\relinline{,}$F_2$\relinline{)}, while $F_1$ \relinline{or} $F_2$ is the same as \relinline{\{}$F_1$\relinline{;}$F_2$\relinline{\}}. 

In \Rel, if we want to condition the evaluation of a \sexpr{RelExpression} on the truth of a formula, we can simply write \mbox{\relinline{(}\sexpr{RelExpression}\relinline{,}} \sexpr{Formula}\relinline{)}. This expression 
returns the result of $\sexpr{RelExpression}$ when $\sexpr{Formula}$ evaluates to true, and $\{\,\}$ otherwise. Indeed, taking the Cartesian product of any relation $R$ with $\{\langle\rangle\}$ is $R$ itself, and taking the Cartesian product of $R$ with the empty relation is empty.

Given the importance of such conditioning in queries (essentially corresponding to the {\tt WHERE} clause in SQL), \Rel has syntactic sugar
\begin{centeredrel*}
    $\sexpr{RelExpression}$ where $\sexpr{Formula}$
\end{centeredrel*}
which is equivalent to 
{\small \relinline{(}\sexpr{RelExpression}\relinline{,}\sexpr{Formula}\relinline{)}}.
Expression (\ref{eq:abstraction-example}) can then be rewritten as \relinline{\{[x,y] : OrderProductQuantity[x] where PaymentOrder(y,x)\}}.

\subsubsection{Linear Algebra}\label{sec:linear-algebra}

Vectors, matrices, and tensors can be modeled with
relations. Vectors are encoded as binary relations:
 the first column holds a position, and the second holds the value at that
position. Matrices are encoded as ternary relations: the first two
columns encode a position (row and column), and the third the value.
Two such examples of encoding a vector of length 4 and a $2\times 2$-matrix are shown below:
\[
  \mathbf{v} = \left(\!\!\begin{tabular}{c} $v_1$ \\ $v_2$ \\ $v_3$ \\ $v_4$ \end{tabular}\!\!\right)
  \ \ \Rightarrow \ \ 
  \begin{array}{c|c} 
  \multicolumn{2}{c}{\mbox{\textit{V}}}\\
  \toprule 
  1 & v_1 \\ 
  2 & v_2 \\
  3 & v_3 \\
  4 & v_4
  \end{array}
  \qquad 
  \mathbf{M} =  \left(\!\!\begin{array}{c@{\hspace{1mm}}c} m_{11} & m_{12} \\ m_{21} & m_{22} \end{array}\!\!\right)
  \ \ \Rightarrow \ \ 
  \begin{array}{c@{\hspace{2mm}}c|c} 
  \multicolumn{3}{c}{\mbox{\textit{M}}}\\
  \toprule
  1 & 1 & m_{11}\\
  1 & 2 & m_{12}\\
  2 & 1 & m_{21}\\ 
  2 & 2 & m_{22}
  \end{array}
\]
The same principle works for tensors: $k$-dimensional tensors are encoded as $(k+1)$-ary relations, where the first $k$ columns encode the tensor coordinates, and the last one the value.

Given two vectors $\mathbf{u}=(u_1,\ldots,u_n)$ and $\mathbf{v}=(v_1,\ldots,v_n)$, their scalar product is
$\mathbf{u}\cdot \mathbf{v} = \sum_k u_iv_i$. \Rel's definition of this mimics the mathematical definition: 
\begin{centeredrel*}
def ScalarProd[{U},{V}] : { sum[[k] : U[k]*V[k]] }
\end{centeredrel*}
Notice the mechanics of this expression: the range of \relinline{k} is guarded by the first columns of \textit{U} and \textit{V}. Subexpression \relinline{[k] : U[k]*V[k]} evaluates to  $\{\langle i, u_iv_i \rangle \mid i \in \{1,\ldots,n\} \}$. The \relinline{sum} aggregate, which computes the sum of the values in the last column of the subexpression, therefore indeed computes $\mathbf{u}\cdot \mathbf{v}$. Crucially, by the definition of \relinline{sum}, it is applied to the \emph{entire relation} $\{\langle i, u_iv_i \rangle \mid i \in \{1,\ldots,n\} \}$, not its projection on the last column.

Indeed, assume that $\mathbf{u} = (4,2)$ and $\mathbf{v}=(3,6)$, which are encoded by $\textit{U} = \{\langle 1,4 \rangle, \langle 2,2 \rangle \}$ and $\textit{V} = \{\langle 1,3\rangle, \langle 2,6\rangle\}$. Then the values of \relinline{k} in \relinline{U[k]} and \relinline{V[k]} are limited to $1$ and $2$. So, \relinline{[k] : U[k]*V[k]} evaluates 
to $\{\langle 1,12 \rangle, \langle 2,12\rangle\}$ and the sum correctly results in 24.

The same approach makes it easy to define matrix
multiplication. Recall that if $\mathbf{M}=\mathbf{A}\cdot\mathbf{B}$,
then its entry $m_{ij}$ in the $i$th row and $j$th column is $\sum_k
a_{ik}\cdot b_{kj}$, as reflected in this \Rel definition:
\begin{centeredrel*}
def MatrixMult[{A},{B},i,j] : { sum[[k] : A[i,k]*B[k,j]] }  
\end{centeredrel*}
Similarly, $\mathbf{A}\cdot\mathbf{v}$ for a matrix $\mathbf{A}$ and a vector $\mathbf{v}$ is defined in \Rel by
\begin{centeredrel*}
def MatrixVector[{A},{V},i] : { sum[[k] : A[i,k]*V[k]] }
\end{centeredrel*}

An advantage of point-free code is that it is more robust under changes of the underlying data. For instance, \relinline{Union[R,S]}  %
and \relinline{MatrixMult[A,B]}, %
work independently of the arities of the relations \relinline{R} and \relinline{S} or the dimensions of the matrices \relinline{A} and \relinline{B}.

\subsection{Graph Library} \label{sec:graphlib}
\paragraph{All Pairs Shortest Paths}
Next, we demonstrate code for the all pairs shortest path problem. The following code expects two relation variables \relinline{V} and \relinline{E}, representing the finite set of vertices and edges (which are pairs of vertices) of some graph.
\begin{centeredrel*}
def APSP({V},{E},x,y,0) : V(x) and V(y) and x = y
def APSP({V},{E},x,y,i) : 
  exists ((z in V) | E(x,z) and APSP[V,E](z,y,i-1)) and 
  not exists ((j in Int) | j < i and APSP[V,E](x,y,j))
\end{centeredrel*}
The first rule states that the shortest path from a node to itself has length~$0$.
The second rule states that the shortest paths from \relinline{x} to \relinline{y} have length $i$ if there exists an out-neighbor \relinline{z} of \relinline{x} such that the shortest paths from \relinline{z} to \relinline{y} have length $i-1$ and we have not already discovered paths shorter than $i$ from \relinline{x} to \relinline{y}. We use relational application in \relinline{APSP[V,E]}, which returns the set of triples $\langle u,v,k \rangle$ such that the shortest paths from $u$ to $v$ has length $k$.
Although we use quantification over the entire relation \relinline{Int},
the \Rel engine will not loop over all integers to compute the query.

We can also define \relinline{APSP} using aggregation and abstraction: 
\begin{centeredrel*}
def APSP({V},{E},x,y,0) : V(x) and V(y) and x = y
def APSP({V},{E},x,y,i) : 
  i = min[(j) : exists((z) | E(x,z) and APSP[V,E](z,y,j-1))]
\end{centeredrel*}

\paragraph{PageRank}
Using the vector and matrix encoding from Section~\ref{sec:linear-algebra}, we can write (a simplified version of) the PageRank algorithm in \Rel. The code also illustrates 
how a \Rel program can perform a number of steps until a stopping condition is met.

\begin{rel}
def dimension[{Matrix}] : max[(k) : Matrix(k,_,_)]
def vector[d,i]  : 1.0/d where range(1,d,1,i)
def abs(x,y) : (x >= 0 and y = x) or (x < 0 and y = -1 * x)
def delta[{Vec1},{Vec2}] : max[[k] : abs[Vec1[k] - Vec2[k]]]
\end{rel}
\begin{rel}
def next[{G},{P}]: {MatrixVector[G,P]}
def stop({G},{P}): {delta[next[G,P],P] < 0.005}
\end{rel}
\begin{rel}
def PageRank[{G}] : 
    {vector[dimension[G]] where empty(PageRank[G])}
def PageRank[{G}] : {next[G,PageRank[G]] 
    where not empty(PageRank[G]) and not stop(G,PageRank[G])}
def PageRank[{G}] : {PageRank[G] where 
    not empty(PageRank[G]) and stop(G,PageRank[G])}
\end{rel}
Here, \relinline{empty} is the emptiness test given by  \relinline{def empty({R}) : not exists( (x...) | R(x...))}, and the predicate \relinline{range(1,d,1,i)} is true for \relinline{i}$= 1,2,\ldots,{}$\relinline{d}.
By calling \relinline{PageRank[M]}, we now do Page\-Rank until the delta between two consecutive iterations is at most 0.005.

\Rel comes with an extensive graph library that provides many common graph algorithms, as well as a library for path finding.

\section{Building Relational Knowledge Graphs in \Rel}
\label{sec:relational-knowledge-graph}

\textcolor{black}{
\emph{Intelligent applications (IAs)} are applications that leverage a variety of AI techniques and reasoners to improve and automate decision making. Building such applications is complex because it requires a combination of disparate technology stacks (e.g., stacks for transactional, analytical, planning, graph, predictive, and prescriptive reasoning), each with its own data management methods and programming paradigm. Combining all these stacks together makes it very difficult and expensive to develop these types of applications.  It is in fact so difficult that there exists  an industry of companies whose sole purpose is to build and maintain such applications.
}

\textcolor{black}{
The emergence of cloud-native relational databases and Data Clouds built with them (e.g., Snowflake~\cite{snowflake}) provides us with an opportunity to rethink the way we build applications.  These data clouds are sufficiently scalable to hold all of the data from a given enterprise. There is a growing appreciation for the value of \emph{semantic layers} as the basis for bringing data and semantics together in these data clouds.   Semantic layers come in two flavors: those based on dimensional models (e.g., Looker, DBT, AtScale) and those based on knowledge graphs (e.g., Palantir, ServiceNow, Blue Yonder).  The former supports analytical tasks and the latter is used for application logic. Semantic layers based on knowledge graphs make it possible to take a \emph{data-centric} approach to application development.  
}

\textcolor{black}{
Relational databases in GNF can be thought of as Relational Knowledge Graphs. Throughout the paper, we argued why \Rel is ideally suited as a language for a new kind of \emph{relational} knowledge graph.  Semantics based on relational knowledge graphs are critical for leveraging LLMs as they are used to provide a description or model of the users' domain of interest.  This 
enables
LLMs to answer questions by reasoning over the knowledge graph, by generating queries that leverage a variety of symbolic and neural reasoners.
}

Indeed, \Rel can be used as the modeling language that expresses database queries, the entire business logic for intelligent applications and, as such, all the concepts needed for a semantic layer, using a single programming paradigm.
The following components allow us to define the concept of a \emph{relational knowledge graph (RKG)}, which provides both the data description and adds semantics to it:
\begin{enumerate}
\item The \emph{relational data model};
\item The \emph{normal form}: \GNFfull\ (\GNF) allowing us to define higher-arity relations in simple terms; and
\item The \emph{language}: \Rel, which enables expressing complex reasoning tasks and application logic in a declarative manner using a single programming paradigm.
\end{enumerate}
In a relational knowledge graph (see \url{https://docs.relational.ai/rel/concepts/relational-knowledge-graphs}) we model the domain as a set of concepts and relationships between them using \GNFfull. In addition, \Rel can define derived concepts and relationships that model the application semantics. 
These can be computed using a mixture of reasoners: rule-based, prescriptive (e.g., SAT solvers, integer or linear programming solvers), and predictive (e.g., GNNs), all of which can be expressed directly in \Rel.

Compared to traditional approaches to knowledge graph management such as RDF coupled with SPARQL and OWL~\cite{sparql}, or property graphs with Cypher/GQL~\cite{sigmod22,cypher,icdt23}, relational knowledge graphs offer several benefits. Firstly, RKGs naturally capture higher-arity relations via GNF, which are more suited to enterprise applications using relational databases, as opposed to the binary relations of the Semantic Web and property graphs. Secondly, RKG support for view definitions using \Rel facilitates accumulating and structuring knowledge --- a feature missing in GQL or the Semantic Web stack. Finally, \Rel uniquely allows us to integrate a variety of symbolic and neural reasoners beyond traditional database querying. Overall, we believe that \Rel makes it possible to define enterprise knowledge graphs based on the relational paradigm, using a single (declarative) paradigm and a relational technology stack.

\section{Rel: Past, Present, and Future} \label{sec:rel:past-present-future}

\paragraph{\hspace{-\parindent}\normalfont\textbf{Before Rel: Influences}}
\label{influences-sec}
To achieve its key design goal of going beyond the sublanguage paradigm, \Rel's design borrows from decades of research in databases and programming languages. %
It follows the paradigm of a small core expanded by user-written libraries 
\cite{Steele99}. Key ideas for \Rel's core come from Boute's functional language based on predicate calculus \cite{Boute} and built around four main constructs: 
identifier, tupling, abstraction, and application.

Much of the computational capabilities of \Rel are based on Datalog-style recursion  with fewer restrictions than traditional textbook Datalog and even SQL's recursive CTE. \Rel's design was influenced by commercial Datalog implementations such as LogiQL of LogicBlox \cite{ArefCGKOPVW15}, Souffl\'e \cite{souffle},  .QL of Semmle (now GitHub) \cite{dotql}, and Dyna \cite{dyna}, with the latter two reflected in \Rel's ability  to handle infinite sets and to combine recursion with numerical computations.  Updating databases via rules applied to control relations is influenced by Dedalus \cite{dedalus} and Statelog \cite{statelog}. 

The use of relation variables in \Rel  borrows some ideas from Data HiLog \cite{datahilog-ross,datahilog-wood}, a function-free fragment of  HiLog \cite{hilog}.
While \Rel does not at this point offer the functionality of storing relation names in databases and using them in queries as in \cite{datahilog-ross}, its parameter-passing is similar to the mechanism described in \cite{datahilog-wood}. %

\Rel influences are not limited to databases. It incorporates works on building bridges between databases and programming languages
including the design of type systems for database queries \cite{BunemanO96,Wong95} and using comprehensions as the basis of query languages \cite{BunemanLSTW94,Wong00}.
Among pure programming language research influences is the ability to overload parameters in abstractions and to apply different definitions depending on types of those parameters, similarly to languages with multiple dispatch \cite{multiple-dispatch} such as Julia \cite{Julia-2017}. In such languages a method can be dynamically dispatched  based on runtime types of its arguments. The use of sets and tuples and  first-order logic notation in queries are also influenced  by SETL, a language for set manipulation that predates relational databases \cite{setl}.

\Rel is also heavily influenced by the fact-based modeling paradigm, particularly by Object-Role Modeling (ORM) \cite{orm-book}. Unlike traditional relational query languages, \Rel is geared towards working with facts about abstract ORM-style attribute-free entities rather than records representing ER-style entities with attributes. Unnamed columns, partial application, last-column aggregation, and  the null-free set semantics are largely consequences of this. 
Also the rich language of integrity constraints---in place of a more classical database schema---reflects the ORM philosophy. 

The ORM-inspired approach to data modeling entails splitting data into many relations and performing many joins. This can be done without sacrificing performance by embracing factorized representations \cite{OlteanuS16} and worst-case optimal joins \cite{NgoPRR18,leapfrog}; the existence of this toolbox enabled many of \Rel's design decisions.

\textsf{Rel}'s approach to solving classical mathematical problems shares similarities with AMPL~\cite{fourer1990ampl},  Alloy~\cite{AlloyAnalyzer2005}, 
and SolverBlox \cite{solverblox}. 
AMPL's integration with solvers for linear, nonlinear, integer, and constraint programming, aligns with \Rel's support for mixed-integer programming and satisfiability \cite{malik-sat,z3}.  Alloy's declarative and relational modeling language and its use of SAT solvers %
resonate with \Rel's focus on modeling business processes and solving predictive and prescriptive analytics problems. SolverBlox incorporated solver abilities into a declarative language by integrating mixed-integer linear programming with 
LogiQL \cite{ArefCGKOPVW15}.

\paragraph{\hspace{-\parindent}\normalfont\textbf{Rel Today}}

\Rel is implemented today as a part of RelationalAI's relational knowledge graph management system. 
This system is available as a co-processor (or native extension) to Snowflake and is available via the Snowflake marketplace. Following the philosophy of  \emph{meeting users where they are} and to facilitate adoption of the language, the first access to \Rel via Snowflake is provided in the form of a Python library that gives users access to many of \Rel's core features described here, as well as  \Rel-written libraries for tasks such as graph analytics. 
This reflects the fact that users' adoption does not happen overnight. As advocated by studies on users' adoption of programming languages \cite{ChasinsGS21,MeyerovichR13}, 
we endeavor to offer them a rewarding initial experience in a familiar environment (e.g., Python and SQL in Snowflake) that encourages them to explore further and gradually get accustomed to all language features.

Many large enterprises are
using \Rel to build applications that include fraud detection, taxation, and supply chain management. The entire business logic for these applications is modeled in \Rel, leveraging its support for programming in the large.
Applications developed in \Rel have run faster, scaled to larger data sets, with drastically smaller (up to 95\%) code bases, when compared to the legacy applications that they replaced.
We see this as a clear sign that the foundational concepts behind \Rel deliver.

\paragraph{\hspace{-\parindent}\normalfont\textbf{The Future of Rel}}
\label{sec:future}
\Rel is an evolving language. Its foundations presented here meet its key design goals of allowing programming in the large and unlocking the full power of the pure relational model. These foundations are stable, but several new features are under development, with partial support for them already in place.

One direction is modeling dynamic behavior via active transition rules that modify the database state. Essentially, this means adopting ideas from Statelog \cite{statelog} and Dedalus \cite{dedalus} that reconcile active and deductive databases by means of incorporating state into Datalog rules. 
Another extension is to fully unlock the potential of syntactic higher-order queries in Data HiLog \cite{datahilog-ross}. In this approach, relations store both data and schema information 
as relation names indicating where relevant data comes from.  \Rel has already taken steps in that direction by allowing  relation variables, and passing relation names as parameters. 
We also plan to enhance the type system of \Rel, providing it with more complete support for ADTs.

A more speculative direction is to revisit an old idea from 1990s: constraint databases and query languages \cite{KKR95,cdb}. Their  clean declarative formalism for programming with infinite sets --- more general than what \Rel currently provides ---  is especially well-suited for spatio-temporal applications. 
Back then 
constraint solvers, which are the backbone of query evaluation for such languages, were not yet sufficiently mature. With their remarkable progress in the past two decades, and \Rel embracing constraint solving, it is possible that these ideas will achieve broader adoption. %

The development of \Rel will also lead to a host of new research problems.
One of them is developing operational semantics of query languages, that takes us one step closer to a precise mathematical formalization.  Such a formalization can be encoded and verified in a proof assistant. It is also more algorithmic than denotational semantics, and thus closer to the style in which query language standards are written.  
Another research problem is extending classical database concepts (relational languages, their equivalence, static analyses) to languages with tuple variables. 
Handling infinite sets in \Rel goes far beyond the realm of traditional query safety in relational languages, and understanding  what is possible requires much additional research. 
Translations between \Rel and  other languages, first and foremost SQL, which must account for \Rel's expressivity (e.g., the just mentioned safety issue) 
are also on the radar.

\begin{acks}
We are deeply indebted to Martin Bravenboer, who is one of the original designers of \Rel and who provided many helpful comments on an early draft of the paper, and to the RelationalAI product and field engineering teams: the compiler team for implementing \Rel, the engine team for implementing its runtime, the infrastructure team for cloud deployment, the user experience team for providing a Python based developer experience, and to the knowledge engineering, data science, and other field engineering teams for being early users of \Rel and providing invaluable feedback that helped improve it.
\end{acks}

\bibliographystyle{ACM-Reference-Format}
\bibliography{main}

\clearpage
\appendix

\setcounter{section}{0}
\renewcommand{\thesection}{\Alph{section}}
\section{Formal Semantics of \Rel Expressions}
\label{sec:maintext-sem}

\begin{figure*}[t]\small%
\newcommand{\exprfootnotemark}{}%
\noindent\begin{minipage}[t]{.4\linewidth}\setlength\abovedisplayskip{0pt}%
$$\begin{array}{@{}rcl@{}}
  \gr{FOBinding} & \gr{::=} &      \gr{Literal} ~\mid~  \gr{ID} ~\mid~ \gr{ID}\!\kw{.}\!\!\kw{.}\!\!\kw{.} ~\mid~ \gr{ID in ID}\\ 
\gr{Binding} & \gr{::=} & \gr{FOBinding}  \ | \ \gr{\kw\{ID\kw\}}
\\
\gr{Expr} & \gr{::=}  &    \gr{Literal} ~\mid~  \gr{ID} ~\mid~  \gr{ID}\!\kw{.}\!\!\kw{.}\!\!\kw{.}  ~\\
     & | &  \gr{\kw(Expr\kw, ...\kw, Expr\kw)}\\
     & | &  \gr{Expr \kw{where} Formula}\\
     & | &  %
            \gr{\kw\{Expr\kw; ...\kw; Expr\kw\}}\\
     & | &  \gr{Formula}\\
     & | &  \gr{\kw[Binding\kw, ...\kw, Binding\kw] \kw: Expr}\\
     & | &  \gr{\kw(Binding\kw, ...\kw, Binding\kw) \kw: Formula}\\
     & | &  \gr{\kw{\{}Expr\kw{\}}\kw[Argument\kw, ...\kw, Argument\kw]}\exprfootnotemark\\
     & | &  \gr{\kw{reduce}\kw{[}\kw{\&\{}Expr\kw{\},}\kw{\&\{}Expr\kw{\}]}}\exprfootnotemark\\  
\end{array}$$
\end{minipage}\hfill
\begin{minipage}[t]{.55\linewidth}\setlength\abovedisplayskip{0pt}%
\begin{equation*}\begin{array}{@{}rcl@{}}
\gr{Formula} & \gr{::=} &      \kw{\{\}} ~\mid~  \kw{\{()\}}\\
     & | &  %
            \gr{\kw{\{}Expr\kw{\}}\kw(Argument\kw, ..., Argument\kw)}\exprfootnotemark\\
     & | &  \gr{\kw{reduce}\kw{(}\kw{\&\{}Expr\kw{\},}\kw{\&\{}Expr\kw{\},?\{}Expr\kw{\})}}\exprfootnotemark\\
     & | &  \gr{Formula \kw{and} Formula}
     ~\mid~
     \gr{Formula \kw{or} Formula}\\
     & | &  \gr{\kw{not} Formula}\\
     & | &  \gr{\kw{exists((}FOBinding\kw, ...\kw, FOBinding\kw) \kw{|} Formula\kw{)}}\\
     & | &  \gr{\kw{forall((}FOBinding\kw, ...\kw, FOBinding\kw) \kw{|} Formula\kw{)}}\\
     & | &  \gr{\kw{(}Formula\kw{)}} \\
\gr{Argument} & \gr{::=} &  \kw{\_} ~\mid~ \kw{\_}\!\kw{.}\!\!\kw{.}\!\!\kw{.} ~\mid~ 
\gr{ID\!\kw{.}\!\!\kw{.}\!\!\kw{.}} ~\mid~ 
\kw{?}\gr{\kw{\{}Expr\kw{\}}} ~\mid~ \kw{\&}\gr{\kw{\{}Expr\kw{\}}}\exprfootnotemark
\\
\gr{RelDef} & \gr{::=} &  \gr{\kw{def} ID %
 \gr{\kw\{Expr\kw\}}}\exprfootnotemark\\
\gr{RelProgram} & \gr{::=} &  \gr{RelDef} ~\mid~ \gr{RelDef RelProgram}\\
 \end{array}
\end{equation*}
\end{minipage}
\vskip-2ex
\caption{Syntax of Rel. {\normalfont Braces around $\kw{\{}\gr{Expr}\kw{\}}$ can often be omitted.}}
\label{fig:syntax}
\end{figure*}

\begin{figure*}\small\newcommand{\nl}{\hspace*{0pt plus 1fill}\linebreak\hspace*{0pt plus 1fill}}
\begin{minipage}[t]{.45\linewidth}
Below, we assume $c\in\const$ and $x,r\in\Ids$.
\begin{itemize}
    \item 
    $\Sem{c}_{\venv} = \{\langle c \rangle\}$
    \item $\Sem{x}_{\venv} = \venv(x)$
    \item $\Sem{\ddd{x}}_{\venv} = \venv(\ddd{x})$
     \item $\Sem{\,\kw{\_}\,}_{\venv} = \{\langle v \rangle \ \mid \ v\in  \Values\}$ 
  \item $\Sem{\,\kw{\_}\!\kw{.}\!\!\kw{.}\!\!\kw{.}\,}_{\venv} = \Tuples_1$
    \item $\sem{\kw{\{}\rexpr_1 \kw{;}\rexpr_2\kw{\}}}_\mu = \sem{\rexpr_1}_\mu \cup \sem{\rexpr_2}_\mu$
    \item $\sem{\kw{(}\rexpr_1 \kw{,}\rexpr_2\kw{)}}_\mu = \sem{\rexpr_1}_\mu \times \sem{\rexpr_2}_\mu$
    \item $\sem{\rexpr \kw{ where } \formula}_\mu= \sem{\rexpr}_\mu \times \sem{\formula}_\mu$  
    \item $\sem{\kw{[\{}x\kw{\}]}\kw{:} \rexpr}_\mu = \{ \langle R \rangle \cdot t \mid R\in \Rels_1,\ \  t \in \sem{\rexpr}_{\mu\lov\{x \mapsto R\}}\}$%
    \item $\sem{\kw{[}c\kw{]}\kw{:} \rexpr}_\renv = \{ \langle c \rangle\cdot t \mid t\in \sem{\rexpr}_{\renv} \}$
\end{itemize}
\end{minipage}\hfill
\begin{minipage}[t]{.55\linewidth}
    \begin{itemize}
    \item $\sem{\kw{[}x\kw{]}\kw{:} \rexpr}_\renv = \{ \langle v\rangle \cdot t \mid v\in \Values, \ \ t\in \sem{\rexpr}_{\renv\lov\{x \mapsto \{\tup{v}\}\}} \}$
    \item $\sem{\kw{[}x\ \kw{in}\ r\kw{]}\kw{:} \rexpr}_\renv = \{ \langle v\rangle \cdot t \mid v \in  \Values, \ \langle v \rangle\in \sem{r}_\renv, \   t\in \sem{\rexpr}_{\renv\lov\{x \mapsto \{\tup{v}\}\}}\}$
    \item $\sem{\kw{[}\ddd{x}\kw{]}\kw{:} \rexpr}_\mu = \{ t \cdot t' \mid t\in \Tuples_1, \ \  t'\in \sem{\rexpr}_{\renv\lov\{\ddd{x} \mapsto \{t\}\}}\}$
    \item $\sem{\kw{(}\bexpr\kw{)}\kw{:} \formula}_\renv = \sem{\kw{[}\bexpr\kw{]}\kw{:} \formula}_\renv$
    \item $\sem{\kw\{\rexpr\kw\}\kw{[\_}\kw{]}}_\mu =   \big\{ t\ \big|\ \langle v \rangle \cdot t \in \sem{\rexpr}_\renv, \ \ v \in \Values \big\}$
    \item $\sem{\kw\{\rexpr\kw\}\kw{[\_}...\kw{]}}_\mu =     \big\{ t\ \big|\ s \cdot t \in \sem{\rexpr}_\renv, \ \ s \in \Tuples_1 \big\}   $
    \item $\sem{\kw\{\rexpr\kw\}\kw{[}\ddd{x}\kw{]}}_\mu =  \big\{ t\ \big|\ s \cdot t \in \sem{\rexpr}_\renv, \ \{s\} =  \sem{\ddd{x}}_\mu \big\} $
\item %
$\sem{\kw\{\rexpr_1\kw\}\kw{[?\{}\rexpr_2\kw{\}]}}_\mu = %
\big\{ t\ \big|\ \langle v \rangle \cdot t \in \sem{\rexpr_1}_\renv, \ v\in \Values, \ \langle v \rangle \in  \sem{\rexpr_2}_\mu \big\}$ 
    
    \item $\sem{\kw\{\rexpr_1\kw\}\kw{[\&\{}\rexpr_2\kw{\}]}}_\mu =
    \big\{ t\ \big|\ \big\langle{\sem{\rexpr_2}_\mu}\big\rangle \cdot t \in \sem{\rexpr_1}_\renv \big\} $
    
    \item $\sem{\kw{reduce}\kw{[}\kw{\&\{}\rexpr_1\kw{\},}\kw{\&\{}\rexpr_2\kw{\}]}}_\renv = v_1 \otimes \cdots \otimes v_{n}$ where  $\sem{\rexpr_1}_\renv$ defines an associative  binary operation $\otimes$ and $\sem{\rexpr_2}_\renv = \{t_1\cdot\tup{v_1},  \dots, t_n\cdot\tup{v_n}\}$\,.
\end{itemize}
\end{minipage}
\vskip-2ex
\caption{Semantics of \Rel expressions. %
}
\label{fig:Exprs}
\end{figure*}

\begin{figure*}\small
\begin{minipage}[t]{\columnwidth}
    \begin{itemize}
    \item $\Sem{\kw{\{()\}}}_{\venv} = \{\emptytup\}$
    \item $\Sem{\kw{\{\}}}_{\venv} = \emptyset$
    \item 
    $\sem{\kw\{\rexpr\kw\}\kw(\argr\kw, \dots\kw, \argr\kw)}_\mu = \sem{\kw\{\rexpr\kw\}\kw[\argr\kw, \dots\kw, \argr\kw]}_\mu \cap \{\emptytup\}$ 
    \item $\sem{\kw\{\rexpr\kw\}\kw{()}}_\mu = \sem{\rexpr}_\mu \cap \{\emptytup\}$
    \end{itemize}
\end{minipage}\hfill
\begin{minipage}[t]{\columnwidth}
\begin{itemize}
    \item $\sem{\formula_1\ \kw{or}\ \formula_2}_\mu = \sem{\formula_1}_\mu \cup \sem{\formula_2}_\mu$
    \item $\sem{\formula_1\ \kw{and}\ \formula_2}_\mu = \sem{\formula_1}_\mu \cap \sem{\formula_2}_\mu$
    \item $\sem{\kw{not}\ \formula}_\mu = \{\emptytup\} - \sem{\formula}_\mu$
    \item $\sem{\kw( \formula \kw)}_\mu= \sem{\formula}_\mu$
\end{itemize}
\end{minipage}
\begin{itemize}    
    \item $\sem{\kw{exists} \kw{(}\ \kw{(}x\kw{)} \ \kw{|}\ \formula\ \kw{)} }_\mu =  \truthvalue{\text{there is } v \in \Values \text{ such that } \sem{\formula}_{\renv\lov\{x \mapsto \{\tup{v}\}\}} = \{\emptytup\}}$ 
    \item $\sem{\kw{exists} \kw{(}\ \kw{(}x\ \kw{in}\ r\kw{)}\ \ \kw{|}\ \formula\ \kw{)} }_\mu = 
        \truthvalue{\text{there is } v\in \Values \text{ such that } \tup{v} \in \sem{r}_\renv \text{ and } \sem{\formula}_{\renv\lov\{x \mapsto \{\tup{v}\}\}} = \{\emptytup\}}$ 
    \item $\sem{\kw{exists} \kw{(}\ \kw{(}\ddd{x}\kw{)} \ \kw{|}\ \formula\ \kw{)} }_\mu = \truthvalue{
        \text{there is } t \in \Tuples_1 \text{ such that } \sem{\formula}_{\renv\lov\{\ddd{x} \mapsto \{t\}\}} = \{\emptytup\}}$     
    \item $\sem{\kw{forall} \kw{(}\ \kw{(}x\kw{)} \ \kw{|}\ \formula\ \kw{)} }_\mu = \truthvalue{\text{for all } v \in \Values \text{ it holds that } \sem{\formula}_{\renv\lov\{x \mapsto \{\tup{v}\}\}} = \{\emptytup\}}$    
    \item $\sem{\kw{forall} \kw{(}\ \kw{(}x\ \kw{in}\ r\kw{)} \ \kw{|}\ \formula\ \kw{)} }_\mu = \truthvalue{\text{for all } v\in \Values \text{ such that } \tup{v}\in \sem{r}_\renv \text{ it holds that } \sem{\formula}_{\renv\lov\{x \mapsto \{\tup{v}\}\}} = \{\emptytup\}}$    
    \item $\sem{\kw{forall} \kw{(}\ \kw{(}\ddd{x}\kw{)} \ \kw{|}\ \formula\ \kw{)} }_\mu = \truthvalue{\text{for all } t \in \Tuples_1 \text{ it holds that } \sem{\formula}_{\renv\lov\{\ddd{x} \mapsto \{t\}\}} = \{\emptytup\}}$    
\item $\sem{\kw{reduce(}\kw{\&\{}\rexpr_1\kw{\},}\kw{\&\{}\rexpr_2\kw{\},}\rexpr_3\kw{)}}_\renv = \truthvalue{\sem{\rexpr_3}_\renv = \sem{\kw{reduce}\kw{[\&\{} \rexpr_1\kw{\},\&\{} \rexpr_2 \kw{\}]}}_\renv}$
\end{itemize}

where $\truthvalue{\textit{Condition}} = \{\emptytup\}$ if \textit{Condition} holds, and $\truthvalue{\textit{Condition}} = \emptyset$ if \textit{Condition} does not hold. 
\label{fig:formulasA}
\vskip-2ex
\caption{Semantics of \Rel Formulas.}
\label{fig:formulas}
\end{figure*}%

\Rel comes equipped with a formal semantics of its language constructs, described in this section. 
Figure~\ref{fig:syntax} gives a (slightly simplified) grammar of the logical core of \Rel. Two most important syntactic constructs are \gr{Expr}, which evaluate to any relation, and \gr{Formula} which evaluate to Boolean values. Note that \relinline{Expr(Argument,...,Argument)} (full application) is a \relinline{Formula} whereas \mbox{\relinline{Expr[Argument,}}\mbox{\relinline{...,Argument]}} (partial application) is not.
\gr{Argument} defines expressions that may be used in  application. Keywords and reserved symbols of the language are written in a bold blue font (like \kw{\{\}} or \relinline{def}) and we use \gr{...} between  separating characters for
a non-empty list. 

Note that \Rel allows some flexibility in the syntax from Fig.~\ref{fig:syntax}. In particular braces around a rule’s body can be omitted if the body is an abstraction.
It is also allowed to write $\sexpr{ID}$ instead of \kw{\{}\sexpr{ID}\kw{\}}, unless it is a rule body or a binding.
In all cases braces can be omitted around expressions which are already in braces.

\smallskip

\emph{Data Model.} We assume a set $\Values$ of constant values, $\Ids$ of \relinline{ID}s from which variables and relation names come, and $\VarIds$ of \mbox{\relinline{ID}\!\kw{.}\!\!\kw{.}\!\!\kw{.}s} %
for names of tuple variables. 
Since we deal with first-order and second-order relations, we need to cleanly differentiate between single values and sets that contain a single element. For example, $\tup{1}$ denotes a unary tuple and $\{\tup{1}\}$ denotes a relation containing a single unary tuple. Formally, we define the following.
\begin{itemize}%
    \item The set $\Tuples_1$ of \emph{first-order tuples} contains all elements $\tup{r_1,\ldots,r_n}$, where $r_i \in \Values$ for all $i \in [n]$.  In particular the empty tuple $\emptytup$ belongs to $\Tuples_1$.
    \item The set $\Rels_1$ of \emph{first-order relations} contains all sets of first-order tuples, that is $\Rels_1=\mathcal{P}(\Tuples_1)$.  Note that $\Rels_1$ contains finite and infinite sets.
    \item 
The set $\Tuples_2$ of \emph{second-order tuples} contains all tuples of the form $\tup{t_1,\dots,t_n}$, where $t_i \in \Values \cup \Rels_1$ for every $i \in [n]$. In particular, $\Tuples_1 \subseteq \Tuples_2$.
    \item The set $\Rels_2$ of \emph{second-order relations} contains all sets of second-order tuples, i.e., $\Rels_2=\mathcal{P}(\Tuples_2) \supseteq \Rels_1$. 
\end{itemize}
For example, $\langle \{ \langle 1,2\rangle, \langle 3,4 \rangle \}, 5 \rangle$ is a tuple from $\Tuples_2$ whose first element is $\{ \langle 1,2\rangle, \langle 3,4 \rangle \} \in \Rels_1$ and the second is $5 \in \Values$.

While \Rel programs can work with higher-order relations (in particular with $\Rels_2$), they can only output first-order relations  (that is, elements of $\Rels_1$). Also recall that a relation (either in $\Rels_1$ or $\Rels_2$) can contain tuples of different arity.

\paragraph{Semantics of Expressions and Formulas.}

The semantics of \Rel expressions (and formulas) is defined with respect to an  \emph{environment} $\venv$, which is a partial mapping that maps identifiers ($\Ids$) to $\Rels_2$ and varargs ($\VarIds$) to singletons in $\Rels_1$ (i.e. relations containing a single element of $\Tuples_1$). Recall here that $\Rels_2$ includes $\Rels_1$, so we can work with both second-order and first-order objects. 

In order to define scoping of variables, we define the $\lov$ operator on environments. Intuitively, $\mu \lov \nu$ extends the environment $\mu$ with the variable assignments in $\nu$. If both $\mu$ and $\nu$ consider a common variable, then $\nu$ takes precedence.
Formally, given two environments $\mu,\nu$ the environment 
$\mu \lov\nu$ has the domain $\dom(\mu)\cup\dom(\nu)$ and
$$(\mu \lov\nu)(x) \df 
\begin{cases} 
\mu(x) & \text{ if }
x\in \dom(\mu) \text{ and } x\not \in \dom(\nu) \,,\\
\nu(x) & \text{ if }x\in \dom(\nu)\,.
\end{cases}$$ 

\paragraph{Normal form} We give semantics for abstraction and quantification with a single binding, application with a single argument, and binary $\kw{;}$ and $\kw{,}$. The semantics for general expressions can be obtained by straightforward syntactic transformations. For example, applications with an arbitrary number of arguments can be rewritten to this normal form as follows.
\begin{multline*}
\kw{\{}\rexpr\kw{\}}\kw{[}Arg_1\kw,Arg_2\kw, \dots\kw, Arg_k\kw{]} \;\equiv\; \\
\kw\{ \dots \kw\{ \kw\{\rexpr\kw{[}Arg_1\kw{]}\kw\}\kw{[}Arg_2\kw] \kw\}\dots \kw\} \kw[Arg_k\kw{]}
\end{multline*}

We use $x$ for identifiers from $\Ids$, $\ddd{x}$ for elements of $\VarIds$ and $c$ for constant values from $\Values$. Additionally, we use $\rexpr$ to refer to the instances of \relinline{Expr} tokens. 
The semantics of a \Rel expression $e$ with respect to an environment $\mu$, denoted $\sem{e}_\renv$, is in Figure~\ref{fig:Exprs}. The semantics of formulas is in Figure~\ref{fig:formulas}. Note that $\sem{e}_\renv$ can be in $\Rels_2$ and possibly infinite.

The semantics of programs defined much like in recursive Datalog
programs~\cite{ABLMP21}, with an added complication of handling higher-order
relations. In broad terms, this is based on the idea of building a
\emph{dependency graph} which allows to track the flow of information, as
explained in Section~\ref{sec:flow}. The information is then propagated in an
iterative fashion until no new facts can be inferred. Details can be found in Appendix~\ref{sec:semantics}.

\paragraph{Disambiguating First- and Second-Order Arguments}
\label{sec:disambituating}

The reader may have noticed annotations \relinline{?} and \relinline{&} in front of arguments that have not been used in any of our examples. These annotations formally disambiguate first- and higher-order arguments. In most production code and real-life examples, such ambiguities never arise and annotations are dropped. 
As an example of an ambiguous application, 
consider two rules for the same relation \relinline{addUp}.
\begin{rel}
def addUp[{A}] : sum[A]
def addUp[x in Int] : x%10 + addUp[(x-x%10)/10] where x >= 0
\end{rel}

The first rule sums up the last column of a relation, and the second sums up the digits of a non-negative integer. 
What should \relverb+addUp[{11;22}]+ evaluate to? The first rule gives tuple $\tup{33}$;  the second rule gives tuples $\tup{2}$ and $\tup{4}$. 
In such cases,  we do not apply both rules to return $\{\tup{2},\tup{4},\tup{33}\}$, but rather require disambiguation by indicating explicitly how the argument should be treated. We use  \relverb+&+ to indicate that we are passing a second-order argument, and \relverb+?+ to indicate  a \emph{first-order} (that is, ordinary) argument; tuple variables are passed as first-order arguments by default. %
That is, \relverb+addUp[?{11;22}]+ evaluates to $\{\tup{2},\tup{4}\}$ and \mbox{\relinline{addUp[&\{11;22\}]}} evaluates to $\{\tup{33}\}$. We can drop \verb+&+ and \verb+?+  if the engine can figure out whether the argument should be passed as first-order or as second-order by examining the definition of the called relation---which is the case in most real-life programs. %
For the expression \relverb+addUp[{11;22}]+ the engine would raise an error, as both first-order and second-order arguments can be passed to \relverb+addUp+, forcing the programmer to disambiguate by using either \verb+&+ or \verb+?+. 

\section{Formal Semantics of \Rel Programs}
\label{sec:semantics}

We define the semantics of a \Rel  program. Recall that in our grammar a \Rel  program is defined as a set of \Rel definitions, which we also call \emph{rules}. A single rule looks as follows:
\begin{centeredrel*}
def ID $\kw{\{} \sexpr{Expr} \kw{\}}$
\end{centeredrel*}
We call \relinline{ID} the \emph{head} of the rule and \sexpr{Expr} the \emph{body} of the rule.

Assume that we have a program $Q$ given as a sequence $r_1
\ r_2 \ \gr{...} r_n$ of rules. 
Let $S_1,\ldots, S_k$ be all relation names mentioned as the heads of rules. Define $G(Q)$, the \emph{dependency graph} of
$Q$, as the graph whose nodes are $S_1,\ldots,S_k$ and where we have
an edge $(S_i,S_j)$ if $S_i$ is used in the body of a rule whose head
is $S_j$; in other words, $S_j$ depends on $S_i$ and the information flows from $S_i$ to $S_j$. 

A set of relation names is a \emph{cluster} if it corresponds to a strongly connected component (SCC) in $G(Q)$. The clusters of $Q$ can be naturally organized into a graph $\widehat G(Q)$: there is an edge from cluster $X$ to cluster $Y$ in $\widehat G(Q)$ if there is an edge in $G(Q)$ from some relation name in $X$ to some relation name in $Y$. Because clusters correspond to SCCs in $G(Q)$, $\widehat G(Q)$ is a directed acyclic graph (DAG). 

To each cluster we can assign a level. A cluster has level $1$ if it has no incoming edges in $\widehat G(Q)$; it has level $\ell+1$ if it has an incoming edge from a cluster of level $\ell$, and all its incoming edges are from clusters of levels at most $\ell$.

Using the levels we partition the relation names $S_1,\dots, S_k$ into \emph{strata}. Stratum $\Sigma_\ell$ is the union of all clusters of level $\ell$. From the definition it follows that there is an $L \geq 1$ such that $\Sigma_\ell \neq \emptyset$ for $\ell \leq L$ and $\Sigma_\ell = \emptyset$ for $\ell > L$. 

\paragraph{Limits of sets.}
Let us fix a set $X$. Consider a sequence $X_0, X_1, X_2, \dots$ of subsets of $X$. The \emph{pointwise limit} of  $X_0, X_1, X_2, \dots$, written as $\lim_{n\to\infty} X_n$, is the set $X_{*} \subseteq X $ such that 
\begin{itemize}
\item $x \in X_{*}$ iff $x$ belongs to almost all sets $X_n$; that is, there is $n_0$ such that $x \in X_n$ for all $n\geq n_0$; and
\item $x\notin X_{*}$ iff $x$ does not belong to almost all sets $X_n$; that is, there is $n_0$ such that $x \notin X_n$ for all $n\geq n_0$. 
\end{itemize}
If some $x$ belongs to infinitely many $X_n$ and does not belong to infinitely many $X_n$, then $\lim_{n\to\infty} X_n$ does not exist.

\paragraph{Limits of functions.}
Consider now a sequence of partial functions $f_0, f_1, f_2, \dots$  from $Y$ to $Z$ such that there is a notion of limit for $Z$; for example,  $Z=\powerset(X)$. We define the \emph{pointwise limit} of $f_0, f_1, f_2, \dots$, written $\lim_{n\to\infty} f_n$ as the partial function $f_{*}$ from $Y$ to $Z$ such that $f_{*}(y) = \lim_{n\to\infty} f_n(y)$ whenever $y\in\dom(f_n)$ for all $n$ and $\lim_{n\to\infty} f_n(y)$ exists, and otherwise $y\notin\dom(f_{*})$.

\paragraph{Fixed points of operators over environments.} The definition of the semantics of a program $Q$ is based on the fixed point of a function $f$ associated to $Q$, that maps environments to environments. In order to define this fixed point we need a notion of the limit of a sequence of environments. 
Recall that environments are functions mapping identifiers to higher-order relations. We work with second order, so environments map identifiers to elements of $\Rels_2$, but everything works the same for any order.  The space $\Rels_2$ is naturally equipped with a notion of limit: we can apply the pointwise limit to sequences of second-order relations because they are sets of second-order tuples. Hence, we can apply the notion of pointwise limit to a sequence of (second-order) environments. 
Now, given a function $f$ that maps (second-order) environments to (second-order) environments, we let $f^{*}$ be the pointwise limit of the sequence of functions $f, f^2, f^3, \dots$, where $f^{n+1}(\renv) = f(f^n(\renv))$.

\paragraph*{Semantics of programs}
Let~$Q$ be a program, and $\relnames$ the set of all relation names used in~$Q$.
We define the semantics of $Q$ with respect to the environment $\mu$ as follows:
\begin{equation*}
  \sem{Q}_{\venv} = f_L^{*} \circ \cdots \circ f_{2}^{*} \circ f_1^{*} \big(\varepsilon\oplus\venv\big)\,,    
\end{equation*}
where~$\varepsilon$ is the environment mapping every relation in~$\relnames$ to the empty set, and 
where $f_i$ is the partial function that maps an environment $\renv'$ with $\relnames\subseteq\dom(\renv')$  
to an environment over~$\dom(\renv')$ defined as 
\begin{equation*}
f_i(\renv')(R)  = \begin{cases}
\bigcup_{\textit{Expression}\ \in\ B_R} \sem{\textit{Expression}}_{\venv'} & \textrm{if } R \in \Sigma_i \qquad  \\
\mu'(R) & \text{otherwise}
\end{cases}
\end{equation*}
\begin{itemize}
\item 
$B_R=\{\textit{Expression}~\mid~ \left(\kw{def}~R~\textit{Expression}\right)\in Q \}$ is the set of bodies of rules defining $R$; 
\item
$f_i^{*}$ is the pointwise limit of $f_i$ as defined above.
\end{itemize}

Note that the body of  each rule $\left(\kw{def}~R~\textit{Expression}\right)$ defining a relation name $R$ in stratum $\Sigma_i$ only uses predicates 
 from $\Sigma_j$ with $j\leq i$. 
Then, the function $f_i$ simulates applying one derivation step to the stratum of $i$ of the program $Q$. The limit $f_i^{*}$ is the result of iterating this operation until saturation.

If we begin with a stratified Datalog program, the semantics above coincides with the usual one, and the iterative computation underlying the semantics terminates. 

We model the database as an environment $\renv_D$ that
maps identifiers of relations stored in the database to their content. Similarly we model \Rel primitives implemented in an external programming language (like \relinline{rel\_primitive\_add}) as an environment $\renv_P$ that
maps identifiers of primitives to their content according to their definition. We assume that $\dom(\mu_D) \cap \dom(\mu_P) = \emptyset$.
Furthermore, we have a standard library, consisting of a set of \Rel rules which we denote by $L$, providing the semantics of library relations. Note that the semantics of some of these relations might be infinite. 
Semantically, we deal with the library by adding  $L$ to each program $Q$; that is, we work with $Q \cup L$, rather than $Q$ alone.
Then, the output of program $Q$ on a database $D$ is 
\begin{equation*}
    \Sem{Q\cup L}_{\venv_D \oplus\venv_P}(\text{\relinline{output}})
\end{equation*} as long as it is a finite relation from $\Rels_1$.

\paragraph*{Note on the semantics of programs} Notice that in our definition the pointwise limit of $f^{*}$ might not be defined for some $\mu$, meaning that the semantics of the program is undefined for this $\mu$. In particular, this means that the semantics of a program might be defined  only on some inputs.

\end{document}